\documentclass[submission, Phys]{SciPost}
\usepackage{subcaption}

\usepackage[utf8]{inputenc} 
\usepackage[T1]{fontenc} 	
\usepackage[english]{babel} 
\usepackage{hyphenat}

\usepackage[bitstream-charter]{mathdesign}
\usepackage{geometry} 		
\usepackage{amsmath} 		
\usepackage{mathtools} 		
\usepackage{float} 			
\usepackage{graphicx} 		
\usepackage{tabularx} 		
\usepackage{booktabs} 		
\usepackage{color, xcolor} 	
\usepackage{pdfpages} 		
\usepackage{extarrows} 		
\usepackage{multirow} 		
\usepackage{multicol} 		
\usepackage{enumitem} 		
\usepackage{xspace} 		
\usepackage{stackrel} 		
\usepackage{tikz} 			
\usepackage{braket} 		
\usepackage{bm} 			
\usepackage{tensor} 		
\usepackage{slashed} 		
\usepackage[group-digits=false]{siunitx} 		
\usepackage{lastpage} 		
\usepackage{cite} 			
\usepackage[normalem]{ulem} 
\usepackage{fontawesome} 	
\usepackage{tocloft} 		
\usepackage{titlesec} 		
\usepackage{doi} 			
\usepackage{hyperref} 		
\usepackage[most]{tcolorbox} 					
\usepackage[nameinlink, capitalize]{cleveref} 	
\usepackage[nottoc, notlot, notlof]{tocbibind} 	
\usepackage[ruled, vlined]{algorithm2e} 		
\usepackage{makecell}
\usepackage[makeroom]{cancel}
\usepackage{feynmf}

\makeatletter
\def\BState{\State\hskip-\ALG@thistlm}
\makeatother

\makeatletter
\@ifundefined{pdfoutput}{}{\DeclareGraphicsRule{*}{mps}{*}{}}
\makeatother

\makeatletter
\DeclareRobustCommand*{\bfseries}{%
   \not@math@alphabet\bfseries\mathbf
   \fontseries\bfdefault\selectfont
   \boldmath
}
\makeatother

\hypersetup{
	pdftitle={XAI4QCDTaggers},
	pdfauthor={Vent et al.},
	colorlinks=true, 			
	linkcolor={red!50!black}, 	
	citecolor={blue!50!black}, 	
	urlcolor={blue!80!black} 	
} 

\DeclareSymbolFont{usualmathcal}{OMS}{cmsy}{m}{n}
\DeclareSymbolFontAlphabet{\mathcal}{usualmathcal}

\SetArgSty{textnormal}
\SetKwComment{Comment}{{\small\#}~}{}
\SetCommentSty{mycommfont}

\setitemize{itemsep=0pt, parsep=0pt} 				
\setenumerate{itemsep=0pt, parsep=0pt} 				
\setitemize{itemsep=2pt,topsep=2pt,parsep=0pt,partopsep=0pt,leftmargin=*}
\setenumerate{itemsep=0pt,topsep=2pt,parsep=0pt,partopsep=0pt,labelindent=3pt,leftmargin=*}
\usepackage{amsmath}
 
\usepackage{amsthm} 		
\theoremstyle{definition}
 
\newcommand{\rej}{\text{Rej}}
\newcommand{\npf}{n_\text{pf}}

\newcommand{\wpf}{w_\text{pf}}
\newcommand{\ptd}{p_TD}

\newcommand{\spid}{S_\text{PID}}
\newcommand{\sfrag}{S_\text{frag}}

\definecolor{red_cb}{HTML}{e41a1c}
\definecolor{blue_cb}{HTML}{377eb8}
\definecolor{green_cb}{HTML}{4daf4a}
\definecolor{purple_cb}{HTML}{984ea3}
\definecolor{orange_cb}{HTML}{ff7f00}

\definecolor{EmeraldGreen}{HTML}{1ea78d}
\definecolor{EnglishRed}{HTML}{b02427}
\hypersetup{colorlinks=true,urlcolor=EmeraldGreen,citecolor=EmeraldGreen,linkcolor=EnglishRed}

\newcommand{\eg}{\text{e.g.}\;}
\newcommand{\ie}{\text{i.e.}\;}

\newcommand{\eqcomma}{\;,} 	
\newcommand{\eqperiod}{\;.} 	

\newcommand{\mwith}{\text{with}}
\newcommand{\mand}{\text{and}}

\newcommand{\Langle}{\bigl\langle}
\newcommand{\Rangle}{\bigr\rangle}

\def\d{\mathrm{d}}

\newcommand\one{\leavevmode\hbox{\small1\normalsize\kern-.33em1}}

\newcommand{\loss}{\mathcal{L}} 	

\newcommand{\pythia}{\texttt{Pythia}\xspace}

\newcommand{\herwig}{\texttt{Herwig}\xspace}
\newcommand{\shap}{\texttt{SHAP}\xspace}

\newcommand{\pysr}{\texttt{PySR}\xspace}

\newcommand{\arXiv}[2][]{%
	\ifthenelse{\equal{#1}{}}%
	{\href{http://arxiv.org/abs/#2}{arXiv:#2}}%
	{\href{http://arxiv.org/abs/#2}{arXiv:#2~[#1]}}}

\def\slashchar#1{\setbox0=\hbox{$#1$}           
   \dimen0=\wd0                                 
   \setbox1=\hbox{/} \dimen1=\wd1               
   \ifdim\dimen0>\dimen1                        
      \rlap{\hbox to \dimen0{\hfil/\hfil}}      
      #1                                        
   \else                                        
      \rlap{\hbox to \dimen1{\hfil$#1$\hfil}}   
      /                                         
   \fi}

\newcommand{\tikznode}[2]{%
\ifmmode%
\tikz[remember picture,baseline=(#1.base),inner sep=0pt] \node (#1) {$#2$};%
\else
\tikz[remember picture,baseline=(#1.base),inner sep=0pt] \node (#1) {#2};%
\fi}

\def\mathswitchr#1{\relax\ifmmode{\mathrm{#1}}\else$\mathrm{#1}$\xspace\fi}
\def\mathswitch#1{\relax\ifmmode#1\else$#1$\xspace\fi}

\newcommand{\PZ}{\mathswitchr Z}

\newcommand{\Pg}{\mathswitchr g}

\newcommand{\Pd}{\mathswitchr d}
\newcommand{\Pu}{\mathswitchr u}
\newcommand{\Ps}{\mathswitchr s}

\newcommand{\jet}{\mathrm{jet}}

\graphicspath{{./figs/}}

\begin{document}


\begin{center}{\Large \textbf{
The Physics Behind ML-based Quark-Gluon Taggers
}}\end{center}
\begin{center}
Sophia Vent\textsuperscript{1,2},
Ramon Winterhalder\textsuperscript{3},
and Tilman Plehn\textsuperscript{1,4}
\end{center}

\begin{center}
{\bf 1} Institut f\"ur Theoretische Physik, Universit\"at Heidelberg, Germany\\
{\bf 2} Dipartimento di Fisica e Astronomia, Universit\'a di Bologna, Italy \\
{\bf 3} TIFLab, Universit\`a degli Studi di Milano \& INFN Sezione di Milano, Italy \\
{\bf 4} Interdisciplinary Center for Scientific Computing (IWR), Universit\"at Heidelberg, Germany
\end{center}

\begin{center}
\today
\end{center}


\section*{Abstract}
{\bf Jet taggers provide an ideal testbed for applying explainability techniques to powerful ML tools. For theoretically and experimentally challenging quark-gluon tagging, we first identify the leading latent features that correlate strongly with physics observables, both in a linear and a non-linear approach. Next, we show how Shapley values can assess feature importance, although the standard implementation assumes independent inputs and can lead to distorted attributions in the presence of correlations. Finally, we use symbolic regression to derive compact formulas to approximate the tagger output.}

\vspace{10pt}
\noindent\rule{\textwidth}{1pt}
\tableofcontents\thispagestyle{fancy}
\noindent\rule{\textwidth}{1pt}
\vspace{10pt}

\clearpage
\section{Introduction}
\label{sec:intro}

Whenever we apply modern machine learning (ML) for fundamental physics~\cite{Plehn:2022ftl}, the same question arises: \emph{Can we identify which physical observables and theoretical structures a trained neural network relies on?}\footnote{We refrain from using this question as the paper title because of Hinchliffe's rule.} 
Understanding the physical basis of the network's decisions gives us insights into their reasoning and strengthen confidence in their theoretical and experimental soundness. We demonstrate how this can be achieved using concepts from explainable AI (XAI), which involves probing the internal representations of trained, high-performance networks.

We take this XAI step by analyzing the inner workings of ParticleNet~\cite{Qu:2019gqs}, a classic architecture for jet tagging. It represents modern ML tools operating on low-level detector inputs, such as jet constituent four-vectors or calorimeter images~\cite{deOliveira:2015xxd}. They have led to significant advances for jet tagging, including advanced transformer implementations~\cite{Qu:2022mxj,He:2023cfc,Wu:2024thh,Brehmer:2024yqw,Spinner:2025prg, Esmail:2025kii}. Specific tasks include quark-gluon tagging~\cite{Komiske:2016rsd,Cheng:2017rdo,Kasieczka:2018lwf,Lee:2019ssx,Lee:2019cad,Kasieczka:2020nyd,Romero:2021qlf,Dreyer:2021hhr,Bright-Thonney:2022xkx,Bogatskiy:2023nnw,Athanasakos:2023fhq,Shen:2023ofd,Dolan:2023abg}, top tagging~\cite{Kasieczka:2017nvn,Macaluso:2018tck,Sahu:2024fzi, Larkoski:2024hfe, Woodward:2024dxb}, W/Z tagging~\cite{Chen:2019uar,CMS:2020poo,Kim:2021gtv, Baron:2023yhw, Colyer:2025ehv, Li:2025tsy}, and bottom/charm identification~\cite{VanStroud:2023ggs,Hassan:2025yhp,ATLAS:2025rbr,ATLAS:2025dkv}. Complementing the classification performance, we look into the trained network and ask:
\begin{enumerate}
\item What features does the network rely on?
\item Do they align with known key observables?
\item Can formulas approximate the network?
\end{enumerate}
  
As a testbed, we choose quark-gluon (QG) tagging~\cite{Nilles:1980ys,Gallicchio:2012ez,Komiske:2018vkc,Larkoski:2019nwj}. While practically extremely promising for many LHC analyses, such as separating signal from background in weak boson fusion or mono-jet searches, QG tagging is theoretically and experimentally tricky. The question of whether a jet originates from a quark or gluon is ill-defined beyond leading order and sensitive to soft and collinear splittings. Furthermore, it strongly depends on the parton shower, hadronization, and detector effects~\cite{Gras:2017jty,Butter:2022xyj}. Very generally, gluon jets radiate more than quark jets because of the color charges, $C_F = 4/3 < C_A = 3 $, so their increased particle multiplicity scales with the ratio of color factors, known as Casimir scaling~\cite{Frye:2017yrw,Larkoski:2014pca}. For additional discriminative power, we want to add more observables with different behavior.

The theoretical and experimental subtleties make QG tagging a particularly compelling case for XAI. Because it lacks a clear-cut ground truth and involves nuanced physics, interpretability is not just a bonus but a necessity. We ultimately envision applying such techniques to networks trained on data, where explainability can drive scientific discovery. Meanwhile, simulation-based studies like ours provide a controlled environment for developing and evaluating XAI tools. While not yet fully established, there is a growing number of physics applications employing promising XAI methods, including Shapley values~\cite{Cornell:2021gut,Grojean:2022mef,Bhattacherjee:2022gjq,Munoz:2023csn,Choudhury:2024crp,Vilardi:2024cwq,Pezoa:2023ulv}, symbolic regression~\cite{Butter:2021rvz,Zhang:2022uqk,cranmer2023_pysr, Soybelman:2024mbv,Morales-Alvarado:2024jrk,AbdusSalam:2024obf,Tsoi:2024pbn,Makke:2025zoy,Bahl:2025jtk}, and other techniques~\cite{Agarwal:2020fpt,Neubauer:2022zpn,Khot:2022aky,Kriesten:2024are, Dimitrova:2025mko,Erdmann:2025xpm,Song:2025pwy}.

In this work, we study the internal representations learned by ParticleNet after training on quark–gluon tagging. Our goal is to investigate how the network uses and combines established physical observables, and whether it encodes additional observables that traditional high-level taggers typically do not exploit. First, we introduce the dataset and describe the ParticleNet architecture used for QG tagging in Sec.~\ref{sec:qg_tagger}. In Sec.~\ref{sec:lat}, we analyze the latent feature space of the network using linear and non-linear dimensionality reduction techniques and investigate how the learned features correlate and share mutual information with known jet observables. In Sec.~\ref{sec:shap}, we perform a Shapley value analysis and discuss its benefits and limitations. Finally, in Sec.~\ref{sec:sr}, we employ symbolic regression (SR) to derive an analytic expression in terms of the leading physical observables that approximates the decision boundary of the ML classifier.

\section{Dataset and classifier network}
\label{sec:qg_tagger}

Distinguishing quark-initiated from gluon-initiated jets is a long-standing challenge in LHC physics. It can enhance precision in Standard Model (SM) measurements and improve sensitivity in searches for Beyond Standard Model (BSM) physics, where signal and background processes often differ in jet flavor composition.
While quark and gluon jets arise from massless QCD splittings, their internal structures differ because of the gluon's larger color charge. This results in higher particle multiplicities and broader radiation patterns.
Beyond these qualitative properties, we investigate the performance of ML-based quark-gluon taggers using precision simulations.

\begin{figure}[b]
\centering
\includegraphics[width=0.30\textwidth]{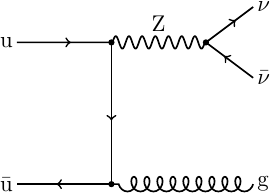}
\hfill
\includegraphics[width=0.30\textwidth]{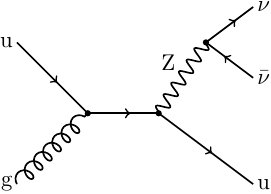}
\hfill
\includegraphics[width=0.30\textwidth]{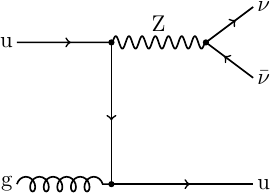}
\caption{Examples of LO Feynman diagrams leading to gluon and quark jets.}
\label{fig:feynman}
\end{figure}

Our primary dataset is generated using \pythia~8.2~\cite{Sjostrand:2014zea,Komiske:2018cqr, komiske_2019_3164691} with default tunes and parton shower settings for the parton-level processes
\begin{align}
q\bar{q} \to \PZ(\to \nu\bar{\nu}) + \Pg \qquad \mand \qquad q\Pg \to \PZ(\to \nu\bar{\nu}) + (\Pu\Pd\Ps) \eqperiod
\label{eq:def_procs}
\end{align}
As the neutrinos remain undetected, these processes provide a clean quark-gluon jet sample, allowing us to investigate any subtle differences in the jet substructure. In Fig.~\ref{fig:feynman}, we show some examples of LO Feynman diagrams for these processes.

Later in our analysis in Sec.~\ref{sec:lat}, we compare results using a similar dataset generated with 
\herwig~7.1~\cite{Bellm:2017bvx, pathak_2019_2664331} to assess robustness across different generators.
Each dataset consists of 2M jets, with up to 100 constituents per jet. We focus on light-flavor jets, and exclude events containing charm or bottom quarks. The jet reconstruction uses the anti-$k_T$ algorithm~\cite{Cacciari:2008gp} with $R = 0.4$, implemented in FastJet~\cite{Cacciari:2011ma}. We select a subset of 600k jets with a training/validation/test split of 400k/100k/100k, each with a 50:50 mixture of quark and gluon jets. For the the labeling of the two processes given in Eq.\eqref{eq:def_procs} we use 
\begin{align}
  \text{jet label} = 
  \begin{cases}
    0 & \PZ(\to \nu\bar{\nu})+g \qquad \; \; \text{(gluon-like)} \\
    1 & \PZ(\to \nu\bar{\nu}) + (\Pu\Pd\Ps) \quad  \text{(quark-like)}
  \end{cases}
\end{align}
Our goal is not to define quark and gluon jets in a theoretically rigorous or generator-inde\-pendent manner, but rather to analyze what structures a network learns when trained on a standard benchmark dataset. For this reason, we adopt the commonly used \pythia~8.2 quark–gluon dataset introduced above~\cite{Komiske:2018cqr,komiske_2019_3164691}, the references dataset for quark–gluon tagging studies. 

While the labeling of quark and gluon jets is inherently ambiguous beyond leading order and subject to generator-specific modeling, this choice is sufficient for our purpose: to interpret the internal representations learned by the network under the same conditions used throughout the literature. Our results should therefore be understood as an analysis of what the network encodes given these standard labels, only approximately related to a proper definition of quark and gluon jets.

\subsubsection*{Low-level classifier}

Historically, quark-gluon taggers have relied on high-level observables motivated by QCD. Modern low-level ML taggers such as ParticleNet~\cite{Qu:2019gqs}, operate on raw information about jet constituents, allowing the network to learn discriminative patterns without any bottleneck. This paradigm shift raises an important question: Are the features learned by such networks consistent with established high-level observables, or do they encode more intricate combinations of the jet constituents?

To address this, we examine the internal representations learned by ParticleNet. We aim to determine whether the network implicitly reconstructs established observables, identifies new combinations of known features, or encodes latent patterns that are difficult to interpret. In subsequent sections, we analyze the structure of the latent space, explore its correlation with physics-motivated observables, and investigate the minimal set of features necessary to preserve full classification performance.

\subsubsection*{ParticleNet}

ParticleNet~\cite{Qu:2019gqs} is a graph convolutional network that processes unordered sets of jet constituents. Each constituent $i$ is represented by a set of low-level features, such as angular distances from the jet axis, momentum, energy, and particle identification (PID). The full set of input features is given by
\begin{align}
    \left \{
    \Delta \eta_i, \Delta \phi_i, \Delta R_i, \;
    \log p_{T,i} ,\log\frac{p_{T,i}}{p_{T,\text{jet}}}, \;
    \log E_i, \log \frac{E_i}{E_\text{jet}}, \;
    \text{PID}_i 
    \right \}\eqperiod
    \label{eq:particle-netobs}
\end{align}
Following the ParticleNet convention, we only consider five different particle categories: electrons, muons, charged hadrons, neutral hadrons, and photons. The electric charge is included in the feature set, consistent with the original ParticleNet design, and we encode the PID by one-hot encoding.

The features in Eq.\eqref{eq:particle-netobs} are passed through a series of edge convolution (EdgeConv) layers. At each layer $l$, the network constructs a dynamic graph by connecting each particle $i$ to its $k$-nearest neighbors $j \in \mathcal{N}(i)$ in the learned feature space. The per-particle feature vector $h_i^{(l)}$ is then updated using learned pairwise interactions:
\begin{align}
h_{i}^{(l+1)} = \sum_{j \in \mathcal{N}(i)} f_\theta^{(l)}\left(h_i^{(l)}, h_j^{(l)} - h_i^{(l)}\right) \eqcomma
\end{align}
where $f_\theta^{(l)}$ denotes a sub-network at layer $l$ with trainable parameters $\theta$. This formulation allows the network to learn local patterns and update the particle features accordingly. At the end, per-particle features are aggregated using average pooling to produce a fixed-size jet representation, which is then passed through a multilayer perceptron (MLP) to output a binary classification probability.

\begin{figure}[t!]
  \includegraphics[width=0.99\linewidth]{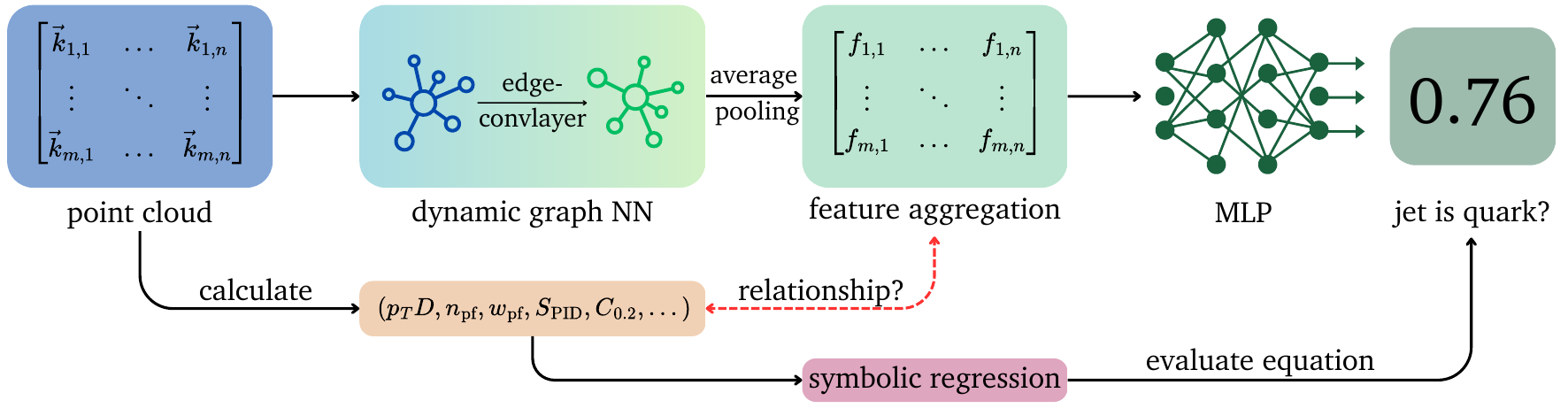}
    \caption{Overview of the ParticleNet architecture and its connection to the explainability techniques explored in our study.}
    \label{fig:ParticleNet}
\end{figure}

Figure~\ref{fig:ParticleNet} illustrates the overall structure of the ParticleNet classifier alongside the inputs and outputs used in our explainability analysis. While the upper path corresponds to the standard inference pipeline described above, the lower path highlights how high-level observables derived from the point cloud can serve as inputs to methods such as symbolic regression or Shapley-based feature attribution. These techniques allow us to probe which physically motivated features the tagger may be implicitly relying on.

We use the compact ParticleNet-Lite variant. It utilizes a single, smaller edge convolution block and outputs a 64-dimensional pooled feature vector per jet. It simplifies the full ParticleNet architecture, which employs two edge convolution blocks and produces a 256-dimensional feature vector.

\section{From latent features to observables}
\label{sec:lat}

ParticleNet-Lite learns a 64-dimensional representation of each jet, but it is not obvious that they are all needed for classification. Compressing the representation is the first step towards explainability. We extract the output of the average pooling layer from ParticleNet-Lite, a 64-dimensional vector summarizing each jet, and study its structure using a linear principal component analysis (PCA) and a latent representation from an autoencoder.

We consider a latent direction \emph{interpretable} if it can be associated with a well-defined physical quantity (\eg multiplicity, fragmentation, charge) rather than only with a complex and entangled mixture of observables. For images, a principal component (PC) might correspond to color or brightness of the jet images. We seek analogous, concise physics semantics and aim to show that the main discriminative directions can be described in these terms.

As a primary alignment score between a latent coordinate $\ell$ (\eg a principal component (PC)) and a jet observable, we use the Pearson correlation coefficient $\rho$. Pearson is bounded within $[-1,1]$ and is scale-invariant, providing a direct and comparable measure of linear correlations across variables. For completeness, we also report mutual information as a non-linear check. However, because it is unbounded, its magnitude is less directly interpretable.

\subsection{Linear dimensionality reduction}

\begin{figure}[b!]
\includegraphics[width=0.495\textwidth]{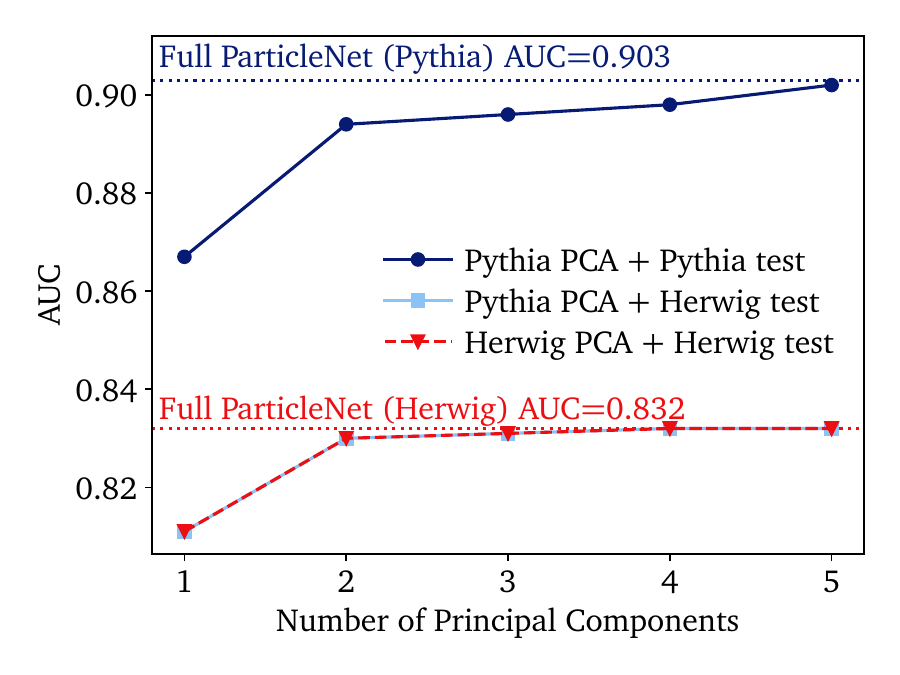}
\includegraphics[width=0.495\textwidth]{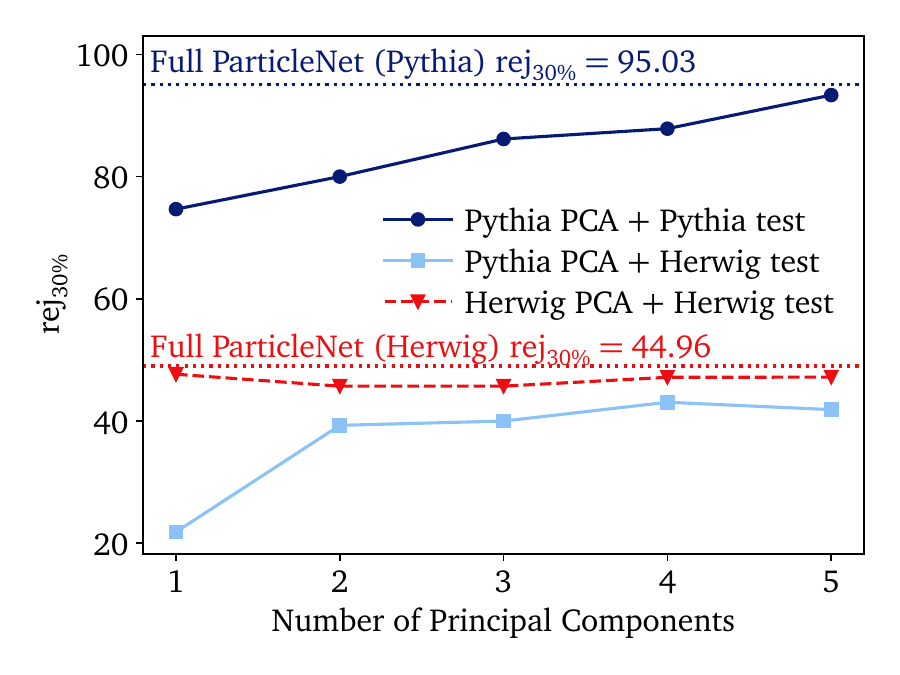}
\caption{Performance of a NN-classifier on the best-performing PC combinations compared to the full ParticleNet-Lite for \herwig and \pythia data sets based on AUC and the rejection rate at 30\% efficiency.}
\label{fig:pca-summary}
\end{figure}

PCA~\cite{Pearson01111901} reduces the dimensionality of data by identifying directions that maximize variance. Let
\begin{align}
 X \in \mathbb{R}^{N \times d} 
 \qquad \text{with} \qquad d=64 
\end{align}
be the features of $N$ jets. After zero-centering each feature, we compute the empirical covariance matrix
\begin{align}
    \Sigma = \frac{1}{N-1}(X-\mu_X)^\top (X-\mu_X)\eqperiod
\end{align}
We then perform an eigen-decomposition
\begin{align}
    \Sigma = V \Sigma_0 V^\top \eqcomma
\end{align}
where $\Sigma_0$ is diagonal, the eigenvalues are called explained variances, \ie  
the leading principal components (PCs) capture directions of maximal variance in the latent space. The matrix $V$ gives the principal directions as eigenvectors, so in the PC basis the jet data is given by
\begin{align}
    Z = (X-\mu) V \eqperiod
\end{align}
To evaluate the impact of the PCs for classification, we train a simple quark-gluon classifier on a set of leading $k$ PCs and determine their AUCs. This tells us how much discriminative power can be retained in lower-dimensional representations. In Fig.~\ref{fig:pca-summary}, we see that the first five principal components are sufficient to recover the ParticleNet-Lite performance, AUC~$=0.902$. The leading three PCs already yield an AUC~$>0.89$.

To ensure resilience, we repeat the analysis using a \herwig dataset and show similar results also in Fig.~\ref{fig:pca-summary}.
Even when the PCA transformation is learned on \pythia jets and applied to \herwig jets, the performance remains comparable. This suggests that the principal directions are relatively universal across generators. Altogether, the performance degrades when using \herwig, relative to \pythia, consistent with previous results~\cite{Butter:2022xyj}.

To understand the compressed latent space learned by ParticleNet, we compare the leading PCs with standard substructure observables like
the particle multiplicity $\npf$, the first radial moment or girth $\wpf$~\cite{Gallicchio:2010dq,Gallicchio:2011xq}, the two-point energy correlation function $C_{\beta}$ for $\beta = 0.2$~\cite{Larkoski:2013eya}, and the width of the $p_T$-distribution of the constituents $\ptd$~\cite{CMS:2012rth},
\begin{align}
     \npf &= \sum_i 1 & \wpf &= \frac{\sum_i p_{T,i} \Delta R_{i,\text{jet}}}{\sum_i p_{T,i}} \notag \\
C_{\beta} &= \frac{\sum_{i<j} p_{T,i} p_{T,j} (\Delta R_{ij})^{\beta}}{\left( \sum_i p_{T,i} \right)^2} & \ptd &= \frac{\sqrt{\sum_i p_{T,i}^2}}{\sum_i p_{T,i}}  \eqperiod
\label{eq:high_level_obs_standard_4}
\end{align}
These observables are chosen as a minimal starting set, since they are commonly used in high-level jet substructure taggers. We then extend this set with additional standard jet observables, such as thrust and higher-order energy correlators, but only report those with the highest correlations. Furthermore, we consider a set of Energy Flow Polynomials (EFPs)~\cite{Komiske:2017aww}, which form an infrared- and collinear-safe, complete linear basis of IRC-safe observables. EFPs summarize a jet by combining momentum fractions $z_i$ with angular separations $\Delta R_{ij}$ in multigraph patterns. A given multigraph is evaluated by multiplying one $z_i$ per vertex and one $\Delta R_{ij}$ per edge, summed over all jet constituents, for example
\begin{align}
\begin{gathered}
\includegraphics[scale=.2]{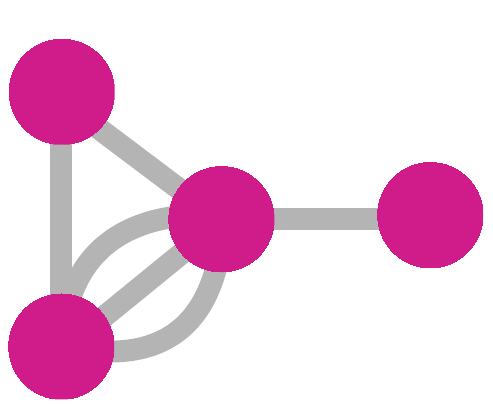}
\end{gathered}
= \sum_{i} \sum_{j} \sum_{k} \sum_{l} z_i z_j z_k z_l\Delta R_{ij}\Delta R_{ik}\Delta R_{jk}^3\Delta R_{kl}\qquad \mwith \qquad z_i = \frac{p_{T,i}}{\sum_i p_{T,i}}\eqperiod
\end{align}
Simple EFPs that are sufficiently described only by the number of vertices $v$ and the number of edges $d$, we denote as 
$\text{EFP}_{v,d}$.
We also consider disconnected multigraphs, defining composite EFPs as
\begin{align}
\text{EFP}_G = \prod_{g \in C(G)}\text{EFP}_g\eqcomma
\end{align}
where $C(G)$ denotes the set of connected components of the multigraph $G$. We include all multigraphs with up to 7 edges, resulting in a overcomplete set of about 1000 EFPs. We then investigate how these observables correlate with the three leading principal components of our trained quark–gluon tagger by evaluating the Pearson correlation coefficient.

\subsubsection*{$\text{PC}_1$: Constituent number and diversity}

\begin{figure}[t]
    \centering
    \includegraphics[width=0.8\linewidth]{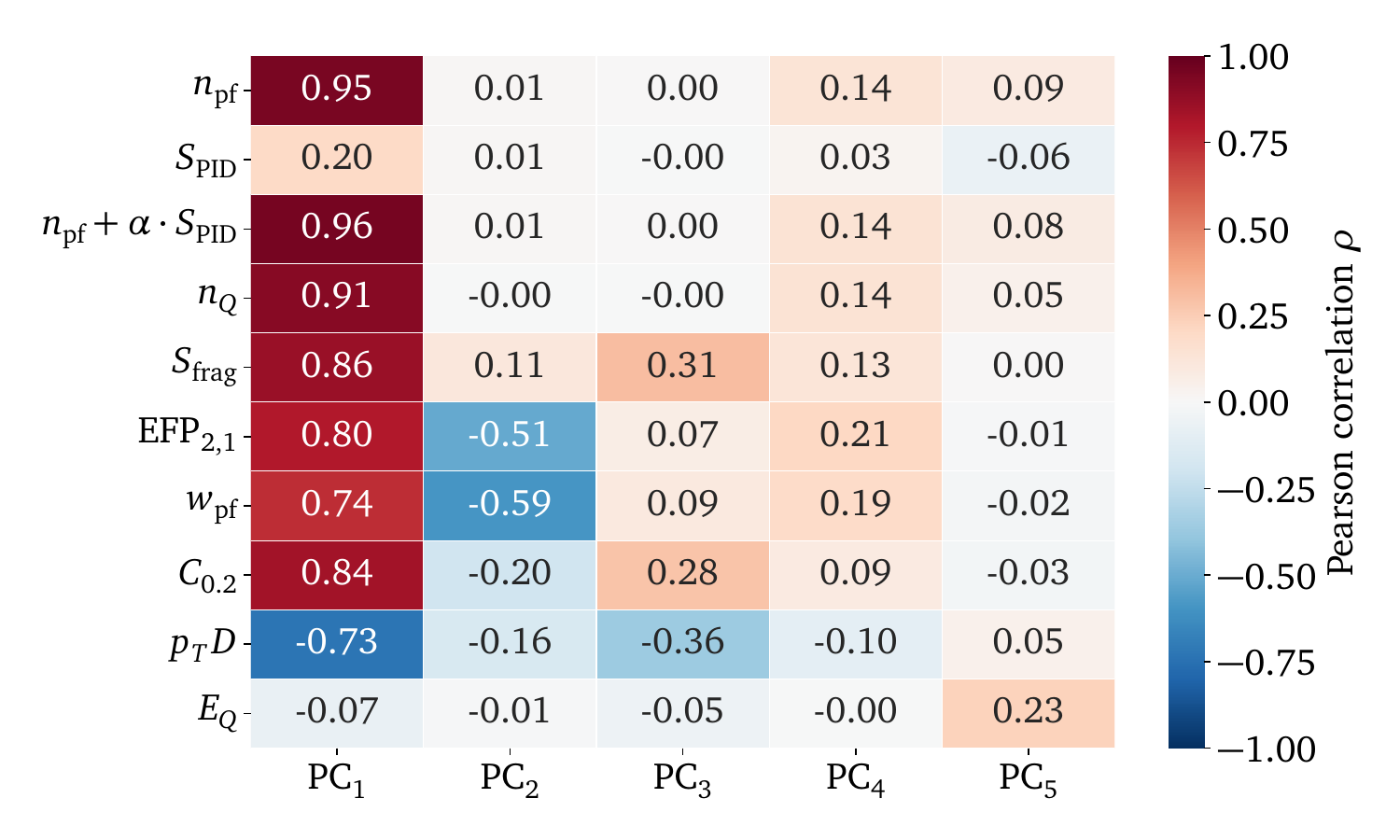}
    \caption{Pearson correlation $\rho$ between the PCs and observables related to multiplicity and particle-type diversity as well as standard observables.}
    \label{fig:PC_1_corr}
\end{figure}

In Figure~\ref{fig:PC_1_corr}, we see that
the first principal component $\text{PC}_1$ is dominated by observables related to the number of particles and their particle nature. Two strongly correlated observables with $\text{PC}_1$ are $\npf$ and the charged multiplicity $n_Q$, defined as the number of charged particles within a jet. In addition, $\text{PC}_1$ is correlated with the PID entropy
\begin{align}
        \spid = -\sum_{\text{type} j} f_j \log f_j \eqcomma
\label{eq:def_spid}
\end{align}
where $f_j$ is the fraction of particles of type $j$. It captures the diversity of particle types in the jet. Gluon jets, which radiate more and produce a broader mix of particles, have larger $\spid$ and multiplicity. Even if the correlation is small, it is relevant since  the linear combination
\begin{align}
    \npf + \alpha \cdot \spid = \npf +  12.3\cdot \spid\eqcomma
\end{align}
achieves a slightly higher correlation with $\text{PC}_1$. The factor $12.3$ was tuned to maximize the correlation with $\text{PC}_1$. This indicates that $\spid$ adds information and is not just correlated with $\text{PC}_1$ through $\npf$.
This means $\text{PC}_1$ reflects both the quantity and diversity of jet constituents.

\subsubsection*{$\text{PC}_2$: Radial energy profile}

\begin{figure}[t]
    \includegraphics[width=0.49\linewidth]{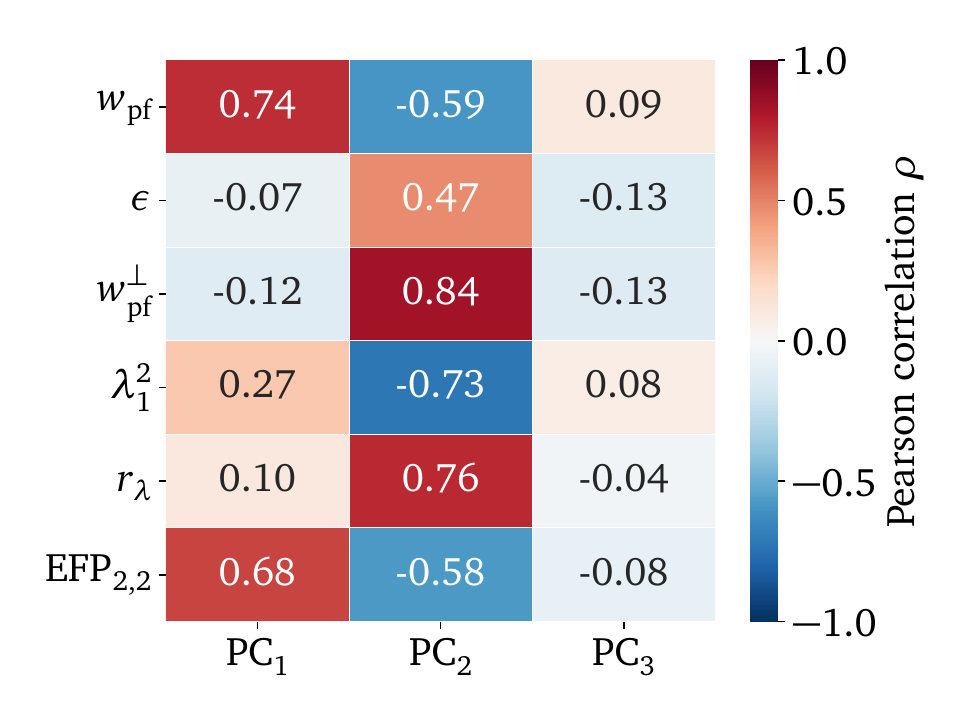}
    \includegraphics[width=0.49\linewidth]{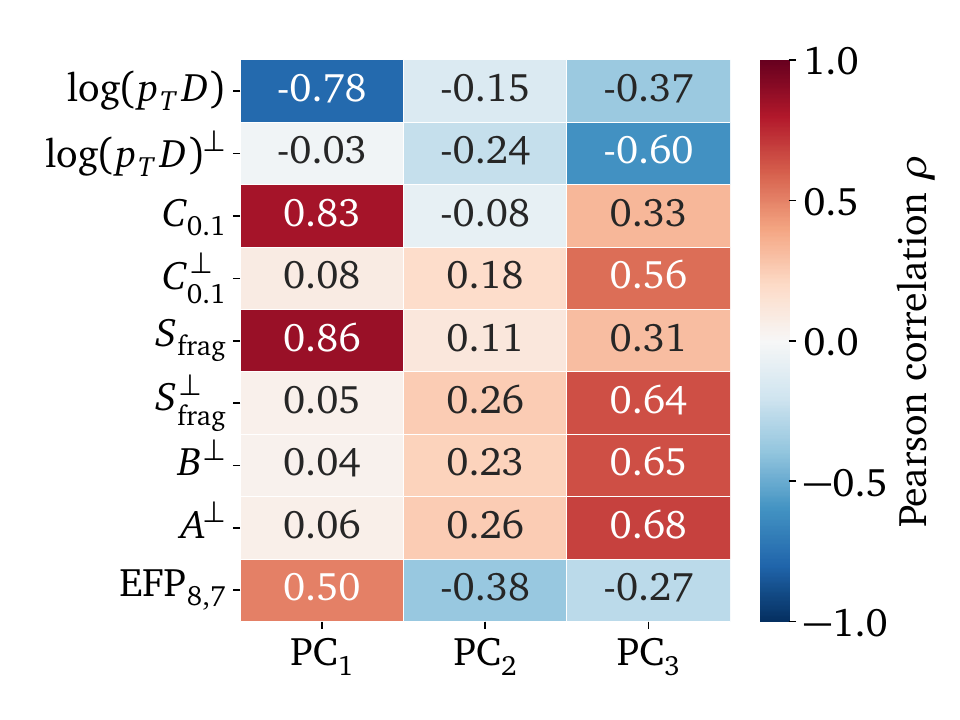}
    \caption{Correlation between the first three principal components and radial jet shape observables(left) and fragmentation and energy dispersion observables (right).}
    \label{fig:PC_2_corr}
\end{figure}

Also in Fig.~\ref{fig:PC_1_corr}, we see that $\text{PC}_2$ captures how energy is distributed radially around the jet axis, independently of multiplicity. It is correlated with several observables sensitive to jet width and shape. An observable only correlated with $\text{PC}_2$ is the ellipticity, defined in terms of the jet inertia tensor~\cite{Brandt:1964sa} in the transverse plane,
\begin{align}
    I^{ij} = \sum_{k\in\jet} p_{T,k} \, \frac{r_k^i r_k^j}{r_k^2} \eqcomma
\end{align}
Here, $r_k^i$ are the components of the transverse position vector of the constituent $k$ relative to the jet axis. The ellipticity is given in terms of $\chi_\text{min}$ and $\chi_\text{max}$ as the eigenvalues of this tensor,
\begin{align}
     \epsilon = \frac{\chi_\text{min}}{\chi_\text{max}} \eqperiod
\end{align}
Lower ellipticity corresponds to more elongated (non-circular) jets. Moreover, $\text{PC}_2$ is strongly correlated with $\wpf$, which is in turn correlated with $\npf$ and $\text{PC}_1$. To exploit this additional direction, we construct the de-correlated combination
\begin{align}
    \wpf^\perp =  \alpha \cdot \npf -\wpf \eqcomma
\end{align}
where $\alpha = 0.0016$ minimizes the linear correlation with $\npf$. It remains sensitive to the jet width while removing the dependence on the multiplicity and therefore removing the correlation to $\text{PC}_1$. The minus sign is chosen since $\wpf$ is negatively correlated with $\text{PC}_2$ and we chose to obtain a positive correlation. In addition, we introduce the generalized angularities~\cite{Gras:2017jty}
\begin{align}
    \lambda_k^\beta = \sum_i z_i^\beta \Delta R_{i,\text{jet}}^k \eqperiod
\end{align}
Among those, $\lambda_1^2$ is strongly correlated with $\text{PC}_2$. In the spirit of energy correlation functions, we can define a ratio between the Les Houches angularity $\lambda_{0.5}^1$~\cite{Andersen:2016qtm, Gras:2017jty} and $\lambda_1^2$
\begin{align}
    r_\lambda = \frac{\lambda_{0.5}^1}{\lambda_1^2} \eqperiod
\label{eq:def_rlambda}
\end{align}
The numerator $\lambda_{0.5}^1$ gives weight to soft emissions at moderate angular scales, typical for gluon jets; the denominator normalizes the broader radial energy profile. This construction has several advantages: (i) it captures the core and the periphery of the jet; (ii) it is dimensionless and robust against global energy rescaling; and (iii) it is naturally decorrelated from the multiplicity as the numerator and denominator share the same structure. Here, $r_\lambda$ serves as an interpretable data-driven proxy for $\text{PC}_2$, not as a perturbative prediction for the full distribution. 
The left panel of Fig.~\ref{fig:PC_2_corr} shows that $\text{PC}_2$ captures the genuine jet shape and radial flow, distinct from $\text{PC}_1$.

\subsubsection*{$\text{PC}_3$: Fragmentation and energy dispersion}

Finally, $\text{PC}_3$ is associated with the way energy is shared among jet constituents, corresponding to the fragmentation pattern. The fragmentation entropy~\cite{Neill:2018uqw} is given by

\begin{align}
    \sfrag = - \sum_i z_i \log z_i\eqcomma
\end{align}
and measures how evenly the transverse momentum is distributed. Quark jets tend to have a lower fragmentation entropy because of their harder fragmentation.

The PCA components are linearly uncorrelated, so $\text{PC}_3$ captures variation in the latent space orthogonal to $\text{PC}_1 \approx \npf$ and $\text{PC}_2 \approx \wpf^\perp$. This suggests that $\text{PC}_3$ encodes physical information that is independent of these two observables and is instead related to fragmentation and energy dispersion.
To test this, we construct candidate observables that may align with $\text{PC}_3$, but remove any linear correlation with $\npf$ and $\wpf^\perp$,
\begin{align}
O^\perp = O - \beta \npf - \gamma \wpf^\perp\eqperiod
\end{align}
where $\beta$ and $\gamma$ are chosen to minimize the linear correlation with $\npf$ and $\wpf^\perp$. This isolates the part of $O$ that corresponds to the latent direction of $\text{PC}_3$. After testing various combinations, the following observables are strongly correlated with $\text{PC}_3$:
\begin{align}
    A^\perp &= \sfrag \frac{C_{0.1}}{C_{0.05}}- 0.03\cdot\npf + 1.95\wpf^\perp \notag \\
     B^\perp &=- C_{0.1} \cdot \log(\ptd) \cdot C_{0.05} -0.014 \npf + 21.32\wpf^\perp \notag \\
   S_{\text{frag}}^\perp &=  S_{\text{frag}} - 0.03 \npf + 0.45\wpf^\perp \notag \\
  C_{0.1}^\perp &= C_{0.1} - 0.0046\npf +0.7701 \wpf^\perp \notag \\
 \log(\ptd)^\perp &= \log(\ptd) +0.0143\cdot \npf- 0.065\cdot \wpf^\perp  \eqperiod
\end{align}
All these observables are sensitive to the distribution of transverse momentum within the jet, characterizing the extent to which the fragmentation pattern is hard or diffuse. In the right panel of Fig.~\ref{fig:PC_2_corr} we see that $\text{PC}_3$ is significantly correlated with these fragmentation-sensitive observables, which means $\text{PC}_3$ capturing aspects of fragmentation dynamics and energy dispersion.

\subsubsection*{$\text{PC}_4$ and $\text{PC}_5$}

Beyond the first three PCs, a clear physical interpretation becomes increasingly challenging. This difficulty arises because many jet observables are highly correlated and tend to span similar directions in feature space. Consequently, the leading PCs capture most of the variance associated with well-understood QCD observables, while the subleading components reflect more subtle structures that may represent combinations of multiple physical effects.

In our analysis, we did not find an individual observable that is strongly or uniquely correlated with $\text{PC}_4$. However, $\text{PC}_5$ shows a notable correlation with the charged energy fraction
\begin{align}
    E_Q =  \frac{E_\text{charged}}{E_\text{jet}} \eqperiod
\end{align}
This suggests that the ParticleNet learns charge-related information in a non-trivial and decorrelated way. Unlike $\text{PC}_{1-3}$, which align closely with standard observables, $\text{PC}_5$ does not map directly onto a single feature but captures a more subtle charge structure of the jet.


Throughout this analysis, we find that standard jet observables outperform individual EFPs when considered as single variables. Because hadronized simulated events possess an explicit infrared cutoff and a finite number of particles, EFPs form a complete linear basis for permutation-invariant observables that depend only on particle momenta. In principle, observables correlated with hadronic multiplicity can be reconstructed from sufficiently high-degree correlators. In practice, however, we use a truncated low-degree EFP basis, in which such information is encoded only indirectly and with limited efficiency. As a result, the leading principal component in our analysis is strongly aligned with a multiplicity-like variation, which is more naturally captured by including the explicit multiplicity observable. This highlights the distinction between numerical completeness on infrared-regulated simulated data and perturbative IRC safety. Observables that depend on non-momentum information, such as particle identity or charge, lie outside the EFP basis altogether.

A further limitation is interpretability. Individual EFPs correspond to graphs, but their detailed physical meaning is not yet associated with clear semantic concepts. Roughly speaking, an EFP with many vertices probes higher-order multi-particle correlations, while a graph with many edges places more weight on angular separations $\Delta R$ and the fine structure of the radiation pattern. For high-degree or composite observables built from many EFPs, this picture becomes increasingly opaque. 
In general, EFPs are well-suited for probing IRC-safe subspaces of jet observables. In this study, we use EFPs mainly as an explicitly IRC-safe reference, not as our main physics observables.

\subsubsection*{Mutual information}

The Pearson correlation $\rho$ only captures the linear relationship between two observables. To quantify non-linear effects, we compute the mutual information (MI). The MI is a measure of the shared information between two random variables. It is usually measured in (Shannon) bits or nats. For continuous variables the mutual information $\text{I(X;Y)}$ of jointly continuous $(X,Y)$ variables is given by
\begin{align}
    \text{I}(X;Y) = \int \d x \d y \, p(x,y) \,\log \frac{p(x,y)}{p(x)p(y)}\eqcomma
    \label{eq:mut_inf}
\end{align}
where $p(x,y)$ is the joint probability density function and $p(x)$ and $p(y)$ are the marginal probability density functions. The MI has no upper bound, making it intrinsically hard to interpret. For continuous variables, the mutual information cannot be evaluated analytically in our setup because the underlying probability densities are not known explicitly, and it must therefore be estimated from finite samples using a $k$-nearest-neighbor estimator as implemented in \texttt{scikit-learn}. As a reference, we also consider the case where $(X,Y)$ has a purely Gaussian dependence with Pearson correlation $\rho$, for which the mutual information is simply given by
\begin{align}
     \text{I}(X;Y)_\text{Gauss} = -\frac{1}{2}\log(1-\rho^2) \eqperiod
    \label{eq:mut_inf_gaus}
\end{align}
By comparing the $k$-nearest-neighbor estimate $I_\text{kNN}(X;Y)$ to this Gaussian baseline, we can test whether the observable exhibits additional non-Gaussian structure beyond what is encoded in the linear correlation.

\begin{table}[t!]
\centering
\begin{small}
\setlength{\tabcolsep}{4pt}
\begin{tabular}{l|S[table-format=1.4(2)]c|S[table-format=1.4(2)]c|S[table-format=1.4(2)]c}
    \toprule
     & \multicolumn{2}{c|}{$\text{PC}_1$} & \multicolumn{2}{c|}{$\text{PC}_2$} & \multicolumn{2}{c}{$\text{PC}_3$} \\
     Observable & {$k$-NN} &  Gauss  &  {$k$-NN}  &  Gauss &  {$k$-NN} & Gauss \\
    \midrule
     $\npf +\alpha\cdot \spid$ & 1.2964(4)& 1.273 & 0.1140(10) & 0.002 & 0.1000(6) & 0.005\\
     $\npf$ & 1.1860(6) & 1.160 & 0.1083(7) & 0.000 & 0.0979(9) & 0.000 \\
     $\text{EFP}_{2,1}$& 0.7578(8) &0.511&0.2683(3)&0.154&0.1019(6)&0.003\\
     $\text{EFP}_{3,3}$ & 0.7804(4) & 0.288 & 0.1974(5) &0.113&0.0830(5) &0.008\\
     \midrule
     $r_\lambda$ & 0.017(2) &0.005 & 0.4772(5) & 0.423 &  0.0114(6) &  0.001\\
     $\wpf$ & 0.6412(8) & 0.398 & 0.3195(7) & 0.210 & 0.1015(6) & 0.005\\
$\wpf^\perp$&0.2123(4)&0.007&0.6520(7)&0.614&0.0355(8)&0.009 \\$\text{EFP}_{2,2}$&0.5937(9)&0.307&0.2482(4)&0.228&0.0720(6)&0.003\\
     $\prod_{i=1}^3\text{EFP}_{2,1}^{(i)}$ &0.755(3)&0.204&0.2709(13)&0.148&0.0991(10)&0.003\\
     \midrule
     $S_{\text{frag}}^\perp$ &0.1077(9) &0.001 &  0.0663(2) & 0.034 & 0.3156(10) & 0.267\\
     $S_{\text{frag}}$ & 0.7481(12) &0.681 &0.1414(11) & 0.006 & 0.1353(2)  &0.051\\
     $C_{0.1}^\perp$ & 0.1705(5) & 0.003 & 0.06345(2) & 0.017 & 0.3037(13) & 0.190\\
     $C_{0.1}$&0.7363(6) & 0.594 & 0.1890(9) & 0.003 & 0.1648(7) & 0.058\\
     $\log(\ptd)^\perp$ & 0.048(2) & 0.000&  0.0475(2)&0.033 &  0.2522(5) & 0.244\\
     $\log(\ptd)$ &0.5094(7) &0.001 &  0.1152(3) & 0.029 &  0.1369(12)& 0.226\\
     $A^\perp$ &0.0905(3) &0.002&0.0589(12) &0.034 &0.3502(7) &0.311 \\
     $B^\perp$&0.0340(10)&0.001& 0.039(2)&0.027&0.2991(14)&0.268\\
$\text{EFP}_{8,7}$&0.2913(4)&0.145&0.0990(7)&0.076&0.0699(2)&0.038\\$\text{EFP}_{2,1}$&0.7578(9)&0.511&0.2692(3)&0.154&0.1019(6)&0.003\\
     \bottomrule
    \end{tabular}
    \end{small}
    \caption{Mutual information in nats between jet observables and principal components using an averaged $k$-nearest-neighbor estimator over various $k$ values and the Gaussian baseline of Eq.\eqref{eq:mut_inf_gaus}.}
    \label{tab:mut_inf}
\end{table}

We evaluate the MI for a number of nearest neighbors $k \in \{3,5,7,12\}$ and average the results, using the standard deviation as an uncertainty estimate. Table~\ref{tab:mut_inf} summarizes the mutual information between various jet observables and the principal components. Alongside the $k$-NN estimate, we show the Gaussian baseline of Eq.\eqref{eq:mut_inf_gaus}, which represents the information content of a purely Gaussian copula with the same Pearson coefficient. When both values agree, the dependence between an observable and a given principal component is adequately described by a Gaussian copula, and we do not resolve additional non-Gaussian structure beyond what is captured by the linear correlation. This is the case for $\npf$  and $\text{PC}_1$, whose $k$-NN and Gaussian mutual information values are very close. In contrast, several observables, such as $\sfrag$, exhibit a mutual information with $\text{PC}_1$ that exceeds the Gaussian reference, indicating additional non-Gaussian features in their joint distribution (\eg non-linear or tail effects) despite already large Pearson correlations.

The Gaussian assumption concerns only the copula of the two variables. Even if the individual marginal distributions are non-Gaussian, their copula can still be Gaussian and lead to the MI in Eq.\eqref{eq:mut_inf_gaus}. A more detailed derivation of this is discussed in App.~\ref{app:mi_copula}. It is also important to stress that absolute MI values across different principal components are hard to compare: each $\text{PC}_i$ has a different entropy, so the maximum achievable MI varies between components. A value that appears small for $\text{PC}_1$ may already saturate the available information for $\text{PC}_2$.

The resulting pattern provides an observable hierarchy that is consistent with the Pearson analysis, and additionally indicates whether a given quantity is effectively encoded through a Gaussian dependence or requires a non-Gaussian structure. A representative example is $\wpf$: its mutual information with $\text{PC}_1$ significantly exceeds the corresponding Gaussian baseline, and a copula fit favors a Student-$t$ distribution over the Gaussian ansatz, indicating enhanced tail dependence between $\wpf$ and $\text{PC}_1$. Once the dependence on $\npf$ is removed, the remaining dependence is well described by a Gaussian copula, and the residual mutual information becomes small. This shows that apparent non-linear relationships can largely be absorbed by leading observables such as $\npf$.

\subsection{Non-linear dimensionality reduction}

\begin{figure}[b!]
    \centering
    \includegraphics[width=0.5\linewidth]{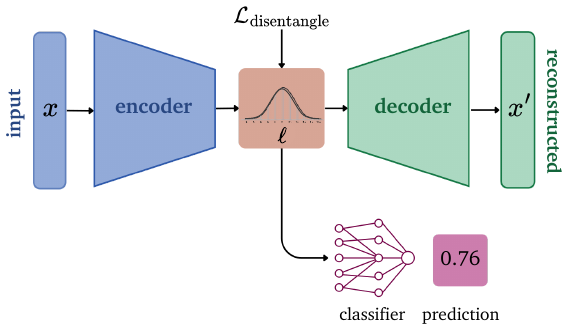}
    \caption{Setup of the DLC. The input is compressed through the
      encoder to a latent space $\ell$, which is enforced to be
      disentangled. The disentangled latent space is then passed to a
      classifier as well as a decoder for reconstruction}
    \label{fig:disentalgeAE}
\end{figure}

Using PCA we have analyzed the latent space of ParticleNet-Lite in a linear approximation. To probe non-linear structures we introduce a Disentangled Latent Classifier (DLC), a network that compresses the 64-dimensional $X$ into a lower-dimensional latent representation, while simultaneously learning to classify quark and gluon jets. The architecture is visualized in Fig.~\ref{fig:disentalgeAE}.
The DLC consists of three components: (i) an encoder that maps each input $x_i \in \mathbb{R}^{64}$, drawn from the dataset $X \in \mathbb{R}^{N \times 64}$, to a latent vector $\ell_i \in \mathbb{R}^d$; (ii) a decoder that reconstructs the original input $\hat{x}_i$ from $\ell_i$; and (iii) a classification head that predicts the jet label $y_i \in \{0,1\}$ from the latent representation. The corresponding loss for $N$ jets is
\begin{align}
\loss = 
\underbrace{\frac{1}{N} \sum_{i=1}^N | x_i - \hat{x}_i |^2}_{\loss_\text{reco}} 
&+ \underbrace{\frac{1}{N} \sum_{i=1}^N \Big[ y_i \log \sigma(\ell_i) + (1 - y_i) \log(1 - \sigma(\ell_i)) \Big]}_{\loss_\text{class}} \notag \\
&+\underbrace{\sum_{j \ne k} \Big[ \text{Cov}(\ell_j, \ell_k) \Big]^2}_{\loss_\text{disentangle}} \eqperiod  
\end{align}
The first term is the MSE between the input vector $x_i$ and its reconstruction $\hat{x}_i$. The second term is the binary cross-entropy, where $\sigma(z_i)$ denotes the predicted probability for jet $i$ in its latent representation. The third term penalizes linear correlations between components of the latent vector by summing the squared off-diagonal elements of the covariance matrix. This loss encourages the network to encode linearly independent features in each latent direction. The covariance penalty is not intended to enforce full (non-linear) independence. Instead, we aim to avoid redundant latent directions that all align with the same dominant factor (\eg $\npf$) or, more generally, with the same reference observable. Since Pearson correlation measures linear dependence, an approximately decorrelated latent space allows us to attribute and rank high-level observables based on their Pearson correlations with individual latent directions. 

By contrast, a mutual information–based disentangling loss is poorly aligned with our objectives. Since our set of observables is intrinsically correlated, minimizing mutual information would favor representations that compress information into a small number of non-linear latent directions rather than distribute it across interpretable components. We therefore rely solely on a linear covariance penalty.

\begin{table}[t]
\centering
\begin{small} \begin{tabular}{c|c c c c}
\toprule
Latent Dim & 1 & 2 & 3 & 4 \\
\midrule
AUC & 0.893(2) & 0.9001(4) & 0.9024(4) & 0.9034(2) \\
$\text{rej}_{30\%}$&72(3) &77(3)& 95(5)& 95(3)\\
$\Delta C$ & 1.8(3) & 0.93(5) & 1.0(16) & 0.9(15)  \\
\bottomrule
\end{tabular} \end{small}
\caption{AUC and calibration for different latent dimensions, averaged over five runs. }
\label{tab:latent_auc}
\end{table}

\begin{figure}[b!]
    \includegraphics[width=0.49\linewidth]{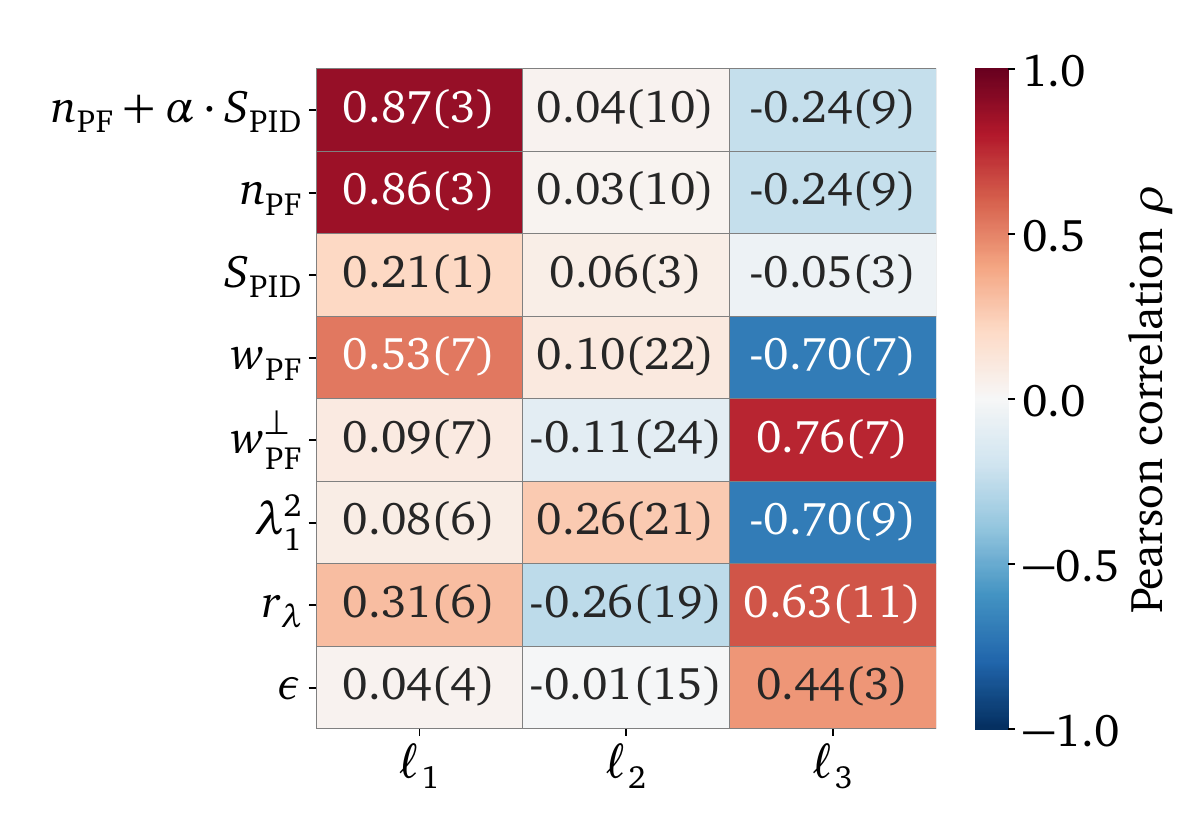}
    \includegraphics[width=0.49\linewidth]{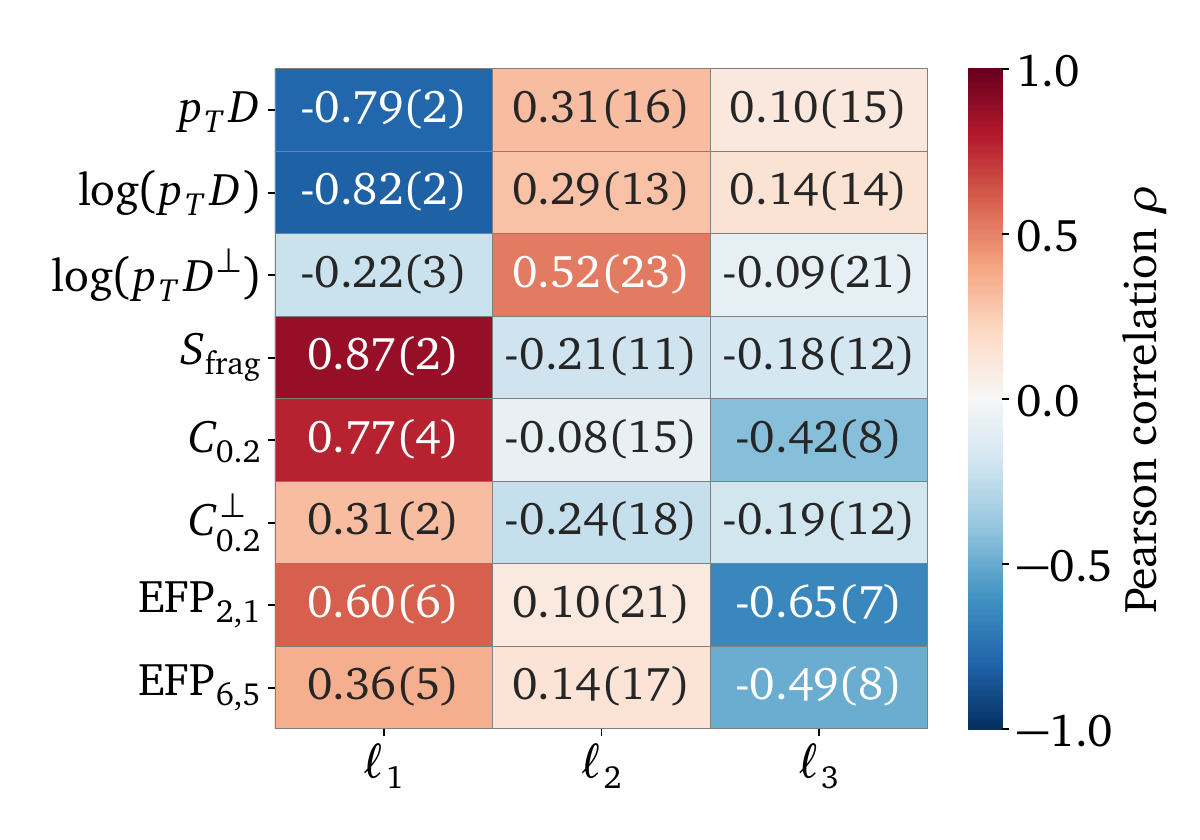}
    \caption{Pearson correlations $\rho$ between physics observables and the linearly disentangled latent spaces $\ell_i$. The uncertainties are computed from 10 independent runs.}
    \label{fig:correlations_latent_spaces}
\end{figure}

To determine the minimal latent dimensionality for successful classification, we train the DLC with different latent space sizes and evaluate the AUC and the calibration. Calibration curves test how well predicted probabilities match observed frequencies. A perfectly calibrated network produces a diagonal curve. To quantify the deviation from the diagonal we use the expected calibration error over $K$ bins of the calibration curve
\begin{align}
\Delta C = 100 \cdot \sum_{k=1}^{K} \frac{N_k}{K} \left| p_k- f_k\right| \eqcomma
\label{eq:def_deltac}
\end{align}
where $N_k$ are the number of events in the $k^\text{th}$ bin, $p_k$ is the average predicted probability in that bin, and $f_k$ is the observed frequency of positive labels. A lower $\Delta C$ means better calibration.  For readability we include a factor of 100.

Table~\ref{tab:latent_auc} shows that a latent space with just three dimensions achieves nearly the same AUC as the full 64-dimensional ParticleNet-Lite output, as well as a decent calibration. In principle, any classifier maps to a scalar output and constructs a one-dimensional representation for discrimination. However, including a reconstruction loss constrains the network to preserve a compressed version of the full feature space, yielding a structured and interpretable latent representation.

In principle, there is no guarantee that the latent directions are the same across different training runs. They can be permuted or have their signs flipped. To test robustness, we train the DLC 10 times with different random seeds. We use the first run as a reference and align the remaining runs to the reference latent directions using the Pearson correlation. If a latent direction $\ell_1$ in the reference run correlates strongly (close to $\pm 1$) with a direction $\ell_2$ in another run, we permute the latent axes accordingly and flip the sign if needed.

After this alignment, we compute Pearson correlations between the jet observables and the latent directions and determine the mean and standard deviation across runs. For 7 out of 9 non-reference runs, a permutation was required, and in total, we applied 13 individual sign flips. This confirms that the order and sign of the latent directions are arbitrary, while their physics interpretation remains stable, as shown in Fig.~\ref{fig:correlations_latent_spaces}. We find a structure very similar to the PCA analysis. The first latent dimension is dominated by $\npf$, the linear combination of $\npf$ and $\spid$ and fragmentation observables, the third by shape observables, and the second by more moderate correlations with fragmentation-related quantities. Overall, the correlations are stable across runs after alignment. Only the correlations with $\ell_2$ vary more, but their correlations with physics observables are also moderate.

While the latent dimensions in DLC are learned to be linearly uncorrelated, their separation is less clean than in PCA. This is because the non-linear transformations can cause overlap in the physical interpretation of different latent directions. For instance, $\log(\ptd)$ is strongly correlated with the first latent direction $\ell_1$, but this correlation disappears once the linear dependence on $\npf$ is removed. This suggests that the correlation is mainly due to the strong dependence of $\log(\ptd)$ on multiplicity, rather than an intrinsic feature of $\ell_1$. While the DLC structure resembles the PCA, the mapping between latent dimensions and physical observables is more general but less direct.

\begin{table}[t]
\centering
\begin{small} \begin{tabular}{ll}
\toprule
Hyperparameter & Value \\
\midrule
Latent dimension $d_\ell$              & 3 \\
Encoder architecture                &  $64 \rightarrow 128 \rightarrow 64 \rightarrow d_\ell$ \\
Decoder architecture                &  $d_\ell \rightarrow 64 \rightarrow 128 \rightarrow 64$ \\
Classifier architecture             & $d_\ell \rightarrow 64 \rightarrow 1$ \\
Activation function                 & ReLU \\
Batch normalization                 & Encoder (128, 64) \\
Dropout (encoder, classifier)       & 0.3 \\
Optimizer                           & Adam \\
Learning rate                       & $10^{-4}$ \\
Batch size                          & 256 \\
Number of epochs                    & 100 \\
Feature scaling                     & StandardScaler (zero mean, unit variance) \\
\bottomrule
\end{tabular} \end{small}
\caption{Hyperparameters of the DLC.}
\label{tab:hyperparams_DLC}
\end{table}

\clearpage
\section{Feature importance from Shapley values}
\label{sec:shap}

Rather than learning and analyzing latent representations, we can train a simple NN-classifier and analyze the feature or observable importance using the SHapley Additive exPlanations (\shap) framework~\cite{NIPS2017_7062}. Shapley values assign a contribution to each input feature for the classifier output $f(x)$, based on cooperative game theory~\cite{RM-670-PR}. For a given feature $i$, the Shapley value $\mathcal{V}_i$ is defined as the average marginal contribution of $i$ across all subsets of the remaining features:
\begin{align}
    \mathcal{V}_i = \sum_{S \subseteq F \setminus \{i\}} \frac{|S|! (|F| - |S| - 1)!}{|F|!} \Big[ f(S \cup \{i\}) - f(S) \Big] \eqperiod
\end{align}
Here, $F$ is the full set of input features (\eg jet observables), and $S$ is a subset of $F$ that does not contain $i$. The term $|S|$ ($|F|$) denotes the number of features in $S$ ($F$), and the sum averages the contribution of feature $i$ over all such subsets. The model output $f(S)$ represents the expected classifier prediction when only the features in $S$ are known. Computing $f(S)$ requires marginalizing over the remaining features $B = F \setminus S$, which is generally intractable. To make this feasible, the model-agnostic kernel \shap makes a simplifying assumption and replaces the conditional distribution $p(x_B \vert x_S)$ with the marginal distribution $p(x_B)$:
\begin{align}
    f(S) = \Langle f(x_S, x_B) \Rangle_{x_B \sim p(x_B \vert x_S)} 
         \approx \Langle f(x_S, x_B) \Rangle_{x_B \sim p(x_B)} \eqperiod
\end{align}
It renders the Shapley analysis numerically feasible, but it can lead to misleading attributions when features are strongly correlated. In such cases, \shap may undervalue features that are informative but share mutual information with others. 

\begin{figure}[b!]
    \includegraphics[width=0.495\linewidth]{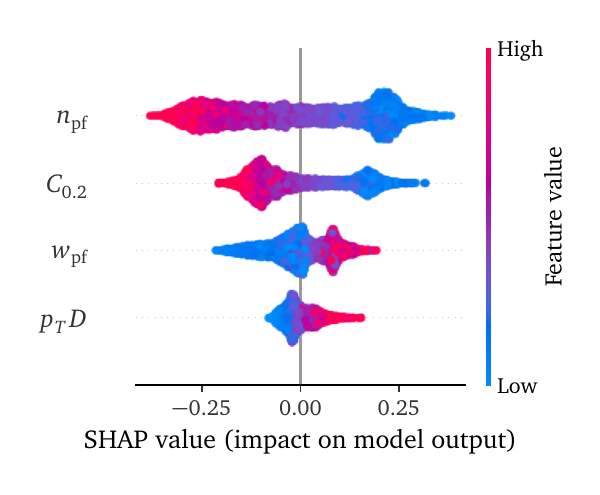}
    \includegraphics[width=0.495\linewidth]{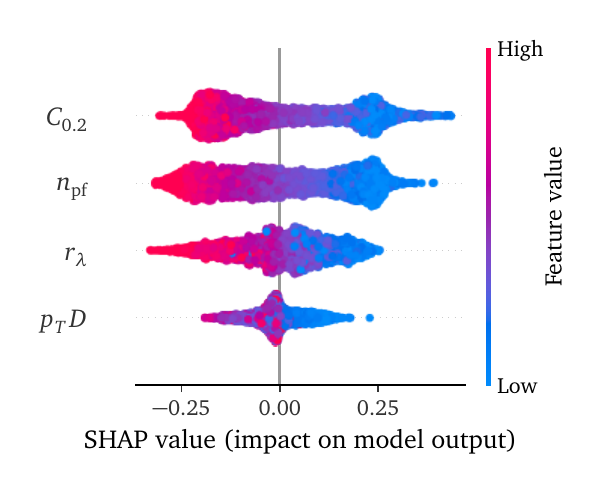}
    \caption{SHAP summary plots for the observables in Eq.\eqref{eq:high_level_obs_standard_4}: baseline set (left) and with $\wpf$ replaced by $r_\lambda$ (right). Observables (features) are ordered by mean absolute SHAP value across events. Each dot is one event; color encodes the observable value for that event (blue = low, red = high). A positive SHAP value indicates stronger contribution towards the quark score.}
    \label{fig:shap_base}
\end{figure}

\begin{table}[t!]
    \centering
    \begin{small} 
    \begin{tabular}{ll|ccc}
    \toprule
    interpretation & Observables & AUC& $\text{rej}_{30\%}$ & $\text{rej}_{50\%}$\\
    \midrule
    standard set &$\npf,\wpf, \ptd, C_{0.2}$ & 0.8626& 71.11&21.53\\
    \midrule
    decorrelated girth &$\npf,r_\lambda, \ptd, C_{0.2}$ &0.8720& 70.00 &23.33\\
    \midrule
    PC1 approx. & $\npf$&0.8406& 53.33& 16.65\\
     &$\npf, \spid$&0.8489& 64.35& 20.32\\
    \midrule
    PC1-2 approx. &$\npf, \spid,\wpf$ & 0.8691&  70.60&21.75 \\
     &$\npf, \spid, \wpf^\perp$ &  0.8690& 72.20 &21.66\\
     &$\npf, \spid, r_\lambda$ & 0.8733&  67.99&21.20\\
    \midrule
    PC1-3 approx. &$\npf, r_\lambda, A^\perp $ &0.8645 & 67.88&19.31\\
     &$\npf,  \spid,r_\lambda, A^\perp $ &0.8774& 73.10&23.45\\
     &$\npf,  \spid,r_\lambda, C_{0.2} $ &0.8778&77.48& 24.78\\
     &$\npf,  \spid,r_\lambda, C_{0.2}, \ptd$ &0.8777&77.24&24.34\\
     &$\npf,  \spid,r_\lambda, C_{0.2}, \ptd, \sfrag$ &0.8825&  79.42&26.80\\
    \midrule
    PC1-5 approx. &$\npf, \spid, r_\lambda, \sfrag, C_{0.2}, \ptd, E_Q$ &  0.8841& 80.89&27.52\\
 
    \bottomrule
    \end{tabular} 
    \end{small}
    \caption{AUC scores for different combinations of three to seven high-level jet observables. }
    \label{tab:performance_different_high_level_obs}
\end{table}

Table~\ref{tab:performance_different_high_level_obs} shows AUC scores for various combinations of high-level observables, to guide our choice of input sets for the \shap analysis.
The left panel of Fig.~\ref{fig:shap_base} shows the \shap summary for the standard tagging observables defined in Eq.\eqref{eq:high_level_obs_standard_4}. 
Positive \shap values indicate that a feature increases the confidence of the network in the quark label, negative values push the prediction towards the gluon label. The features $\npf$ and $C_{0.2}$ behave as expected: low particle multiplicity and a small energy correlation suggest a quark jet. We also see that $\ptd$ contributes little to the classification. 

The feature $\wpf$ in the same panel displays a counter-intuitive pattern: jets with low $\wpf$, typically indicative of narrow, quark-like jets, receive negative \shap values related to a gluon classification. This is not a failure of the classifier, but a limitation of the \shap attribution mechanism.
Low $\wpf$ occurs in both, quark jets (with low $\npf$) and some gluon jets (with high $\npf$), due to their correlation.
The classifier correctly learns that low $\wpf$ combined with high $\npf$ is characteristic of gluon jets.
However, \shap evaluates the contribution of $\wpf$ by marginalizing over $\npf$ and other features, assuming independence and thereby ignoring their joint structure.
As a result, \shap assigns a negative contribution to $\wpf$ even when, conditional on $\npf$, it should favor a quark classification.

To address this mis-attribution, we replace $\wpf$ with $r_\lambda$, the decorrelated alternative introduced in Eq.\eqref{eq:def_rlambda}, as part of a minimal input set,
\begin{align}
    \left\{ \;  \npf, {\color{red!80!black}\wpf}, \ptd, C_{0.2} \; \right\} \quad \longrightarrow \quad 
    \left\{ \;  \npf, {\color{blue!60!black}r_\lambda}, \ptd, C_{0.2}  \; \right\} \; .
\end{align}
In the right panel of Fig.~\ref{fig:shap_base} the features $r_\lambda$, $\npf$ and $C_{0.2}$,
now show a straightforward interpretation. The remaining issue is again methodological: While replacing $\wpf$ by $r_\lambda$ removes the strongest width–multiplicity, strong correlations among the remaining observables persist. In particular, $\ptd$ is strongly anti-correlated with $\npf$. Consequently, the marginalization used by \shap breaks this dependence and the resulting $\ptd$ attributions again reflect a counterfactual variation of one observable while the correlated one is effectively unconstrained. This is the same mechanism that produced the misleading $\wpf$ attribution in the left panel of Fig.~\ref{fig:shap_base}. 

\begin{figure}[t]
    \includegraphics[width=0.495\linewidth]{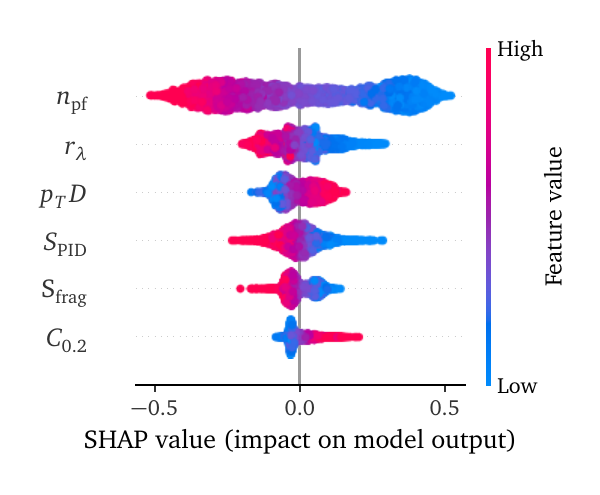}
    \includegraphics[width=0.495\linewidth]{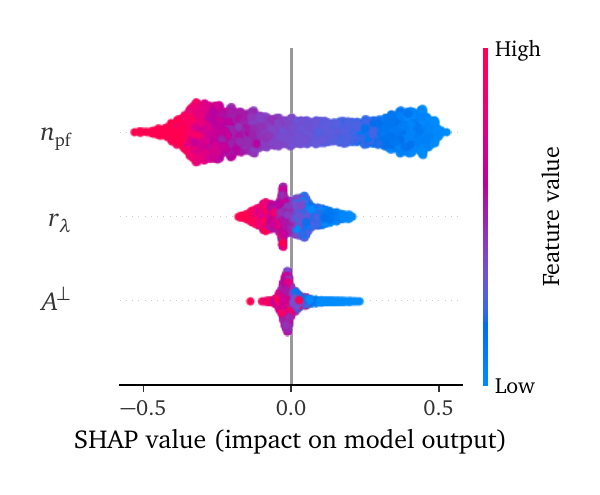}
    \caption{\shap summaries for set of observables approximating the first three PCA components. Left: strong correlations among features distort \shap attributions, particularly for $C_{0.2}$ where the importance direction is counterintuitive. Right: \shap values for a decorrelated feature set approximating the first three PCA components. The importance of each feature aligns with the ranking of principal components.}
    \label{fig:shap_PC_1-3}
\end{figure}

The problem becomes more visible once we increase the number of correlated inputs. The left panel of Fig.~\ref{fig:shap_PC_1-3} shows \shap values for a classifier trained on observables intended to approximate the first three principal components from Sec.~\ref{sec:lat}. Here $C_{0.2}$ displays the opposite qualitative behaviour compared to Fig.~\ref{fig:shap_base}. This is not a change in the physical information carried by $C_{0.2}$. It is a consequence of correlated inputs: \shap distributes shared information among features in a basis-dependent way, which can change both the apparent ranking and the apparent sign structure. This also clarifies why $\npf$ and $C_{0.2}$ swap roles between Fig.~\ref{fig:shap_base} and Fig.~\ref{fig:shap_PC_1-3}. Our PCA analysis, together with long-established expectations from quark–gluon tagging studies, indicates that the multiplicity direction is dominant. The fact that $C_{0.2}$ appears first in Fig.~\ref{fig:shap_base} should therefore be interpreted as a correlation artifact, \ie, multiplicity information is partially attributed to a correlated proxy. When additional correlated observables are introduced, this bookkeeping changes and more of the shared contribution is assigned back to $\npf$, so that the ordering becomes closer to the physically expected one. The absolute magnitudes of the \shap values, however, remain basis-dependent in the correlated setting and should not be overinterpreted. To obtain stable and physically meaningful attributions, we therefore work in an approximately decorrelated basis guided by the PCA interpretation and use
\begin{align}
\{ \npf, r_\lambda, A^\perp \}.
\end{align}
We could have picked other observables for the third component, but we use $A^\perp$ since it combines energy correlations and fragmentation entropy. The \shap plot for this set is shown in the right panel of Fig.~\ref{fig:shap_PC_1-3}, and it aligns with our earlier PCA findings. $\npf$ is clearly the leading feature, with $r_\lambda$ and $A^\perp$ adding extra discrimination.

By default, \shap ranks features by their average absolute contribution. When the inputs are properly decorrelated, this ranking generally matches both intuition and our PCA results, making it easier to interpret. However, with correlated features, \shap gives misleading attributions, even if the overall classifier still works fine. This means we need to carefully prepare inputs, using decorrelated features or a PCA basis, to get \shap explanations that actually reflect the physics. \shap is still a powerful tool, but we have to be mindful of these subtleties when applying it to jet observables.

\clearpage
\section{Symbolic regression}
\label{sec:sr}

Having analyzed the internal structure of the trained network using principal components and Shapley values, we now ask directly: Can the classifier output be approximated by a formula built from high-level observables? This question leads directly to the language of theoretical physics, \ie formulas and equations. In principle, neural networks can be approximated by formulas, and the extremely strong regularization of formulas can be helpful in cases of (too) little training data~\cite{Bahl:2025jtk}. Instead of reasoning about latent vectors or weight matrices, we aim to represent the trained ParticleNet as a formula, capturing its dependencies on subjet observables. Our goal is to express the machine-learned decision boundaries in terms of known physical quantities.

We perform symbolic regression using the \pysr framework~\cite{cranmer2023_pysr}, which searches for formulas by evolving a population of candidate formulas through a genetic algorithm. Each candidate is represented as a tree, constructed from a predefined set of mathematical operations. Each node in the tree contributes to the complexity. For example, the equation
\begin{align}
    3x + a
\end{align}
has a complexity of five, three for the parameters and two for the operations. The algorithm balances two competing objectives, accuracy and simplicity. This makes \pysr particularly well suited for discovering compact formulas that approximate the network output. 

\subsubsection*{Setup and method}

We first select a set of observables based on their performance and interpretability, as discussed in the previous sections. These include particle multiplicity, radial energy distribution, fragmentation entropy, momentum balance, the two-point correlation function, the charged energy fraction, and the PID entropy,
\begin{align}
 \left\{ \npf, \spid, r_\lambda, \sfrag, \ptd, C_{0.2}, E_Q 
 \right\}\eqperiod
\label{eq:def7}
\end{align}
For each input observable (or combination), we first train a simple neural network classifier that uses only those observables. Its output defines the target for the symbolic regression. This isolates the contribution of the chosen observables and ensures that the formulas approximate a realistic, learnable decision function.

For symbolic regression, we use \pysr with a fixed set of operators including addition, multiplication, division, powers, and a small number of non-linear activation functions such as $\tanh(x)$. 
For single-observable fits, we allow a maximum complexity of 10; for two-observable combinations, we increase the limit to 22. The formulas are evaluated based on three criteria that balance precision and interpretability:
\begin{enumerate}
\item the area under the ROC curve (AUC);
\item the background rejection at 30\% quark efficiency; and 
\item the calibration metric $\Delta C$ defined in Eq.\eqref{eq:def_deltac}.
\end{enumerate}
%

\begin{table}[b!]
    \centering
    \begin{small} \begin{tabular}{lc}
        \toprule
        Hyperparameter & Value \\
        \midrule
        iterations & 5000\\
        cycles per iteration & 800\\
        binary operators & $+,\,-, \,\times\eqcomma\,\div$\\
        unary operators & $x^2,x^3, \sqrt{x}, \tanh x $\\
        populations &70\\
        population size &40 \\
        procs &32\\
        batching & True\\
        maxsize & 10 (1D), 22 (2D), 40 (7D)\\
        precision & 32\\
        turbo & True\\
        warmup maxsize by & 0.05\\
    \bottomrule
    \end{tabular} \end{small}
    \caption{Hyperparameter settings for \pysr.}
    \label{tab:placeholder}
\end{table}

\subsection{One-dimensional regression}

We begin by applying symbolic regression to individual high-level observables, to see whether the tagger’s decision surface, restricted to a single input observable, can be captured by a simple formula. The one-dimensional regressions serve mainly as a controlled validation of the symbolic-regression setup before moving to multi-observable fits.The one-dimensional regressions serve mainly as a controlled validation of the symbolic-regression setup before moving to multi-observable fits. For each observable $\mathcal{O}$, we train a neural network using only $\mathcal{O}$ as input and record its predicted quark probability $y_{\mathcal{O}}$. Symbolic regression is then used to approximate
\begin{align}
    f(\mathcal{O}) \approx y_{\mathcal{O}}\eqperiod
\end{align}
To understand the role of functional choices, we first focus on the particle multiplicity $\npf$ as the most discriminative observable for quark-gluon tagging. 

\subsubsection*{Activation functions}

\begin{figure}[t]
    \centering
    \includegraphics[width=0.6\linewidth]{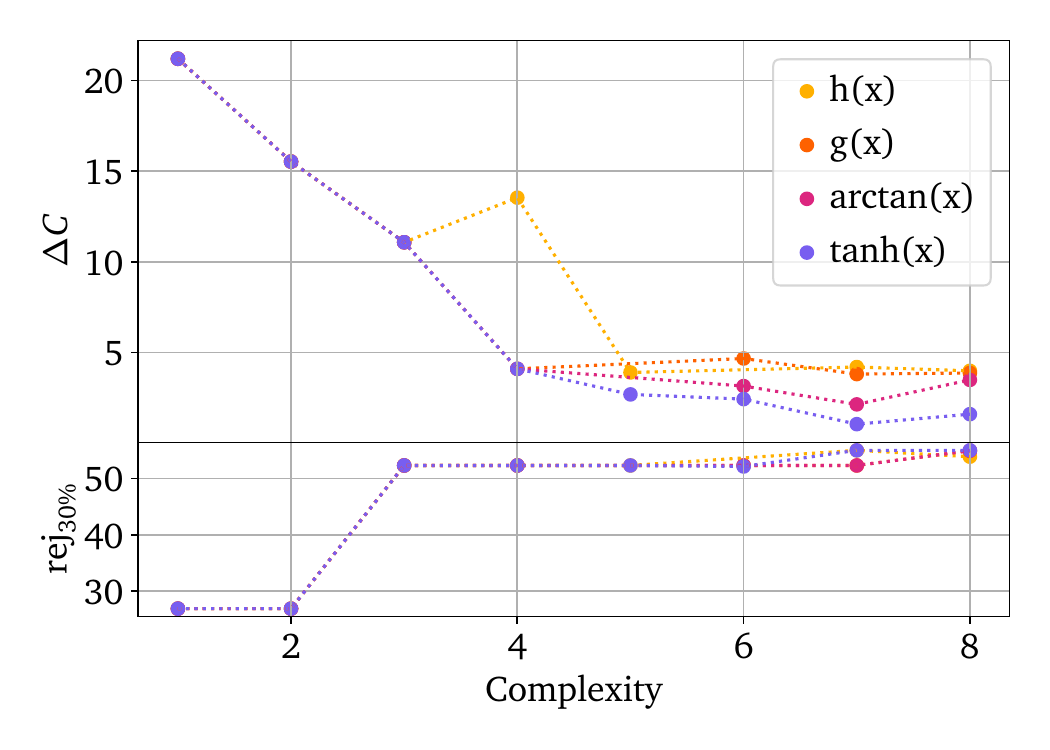}
    \caption{Comparison of activation functions in symbolic regression using $\npf$. The calibration error $\Delta C$ and background rejection at 30\% quark efficiency are shown as a function of formula complexity.}
    \label{fig:test_funcs}
\end{figure}

We first study how the choice of non-linear activation functions affects the learned formulas. Since the tagger outputs are bounded between 0 and 1, it is natural to include bounded non-linearities that can model this range effectively. At the same time, we want to keep the formulas as compact and interpretable as possible. We test standard options like $\tanh(x)$ and $\arctan(x)$, as well as two custom choices
\begin{align}
h(x) = \frac{-x}{\sqrt{1 + x^2}}
\qquad \mand \qquad 
g(x) = \frac{1}{\pi} \arctan(x) + 0.5 \eqperiod
\label{eq:bounded_funcs}
\end{align}
In principle, \pysr can build complex activation functions such as the sigmoid from elementary functions, but this would require a complexity of 19. Hence, explicitly providing more complex activation functions significantly boosts the performance. However, there are two reasons for limiting the symbolic algorithm to a single non-linear function. First, \pysr develops a bias toward an activation function appearing early in its formula search. Once a function like $\tanh(x)$ appears, the evolutionary search typically continues to build on it and ignores the better performance of alternative functions. Second, fixing the activation function reduces the complexity and prevents overly convoluted structures. 

To explore this in a controlled setting, we perform a scan using only $\npf$ as input, and vary the allowed complexity from 1 to 10. For each function, we track the calibration error $\Delta C$ and the background rejection at 30\% signal efficiency. Figure~\ref{fig:test_funcs} summarizes the results.

The different activation functions perform similarly in terms of AUC, but $\tanh(x)$ consistently produces better-calibrated outputs and leads to shorter, cleaner formulas. Based on this, we adopt $\tanh(x)$ as the default non-linearity for all remaining regression tasks. Note that we also considered the sigmoid function in earlier tests, but it was excluded from the final analysis due to its inferior performance.

\subsubsection*{Monotonicity and constant AUC}

\begin{table}
\centering
\begin{small}
\begin{tabular}{c p{7cm}|S[table-format=1.5] S[table-format=1.3] S[table-format=2.2] S[table-format=2.2]}
\toprule
 Complexity& Formula & {Loss}& {AUC} &{$\rej_{30\%}$}& {$\Delta C$} \\
 \midrule
1&0.5&0.083& 0.5&3.33& {-}\\
3&$\frac{17.7}{\npf}$&0.029&0.839&52.32&11.34\\
4&$\tanh \frac{22.3}{\npf} $&0.0169&0.839&52.32&12.23\\
5&$\tanh \frac{850.9}{\npf^{2}} $&0.0009&0.839&52.32&3.43\\
6&$\tanh^{6} \frac{57.447636}{\npf} $&0.0006&0.839&52.32&2.04\\
7&$\tanh \frac{44.19}{\left(0.084 \cdot \npf + 1\right)^{3}} $&0.00026&0.839&52.32&1.64\\
8&$\tanh^{3}{\left(681.83\cdot \left(0.014 + \frac{1}{\npf}\right)^{2} \right)}$&0.00025&0.839&52.32&1.58\\
9&$0.94 \cdot \tanh{\left(21036 \cdot\left(0.005 + \frac{1}{\npf}\right)^{3} \right)}$&0.00004&0.839&52.32&1.08\\
\bottomrule
\end{tabular}
\end{small}
\caption{1D symbolic regression results for $\npf$ only.}
\label{tab:npf_SR}
\end{table}

Looking at  Table~\ref{tab:npf_SR}, we see how the formula evolves with complexity. At low complexity, the network starts with a simple inverse scaling, $\sim 1/\npf$, capturing the trend that higher multiplicities are associated with gluon jets. As complexity increases, \pysr sharpens this behavior by adding non-linear functions like $\tanh$ and raising them to higher powers. These refinements do not change the overall monotonic trend, but they improve the match to the classifier output. From complexity 7, additional gains come mainly from fine-tuning the shape. The formulas remain compact and interpretable, with increasing agreement with the network output.

\begin{figure}[b!]
    \centering
    \includegraphics[width=0.49\linewidth]{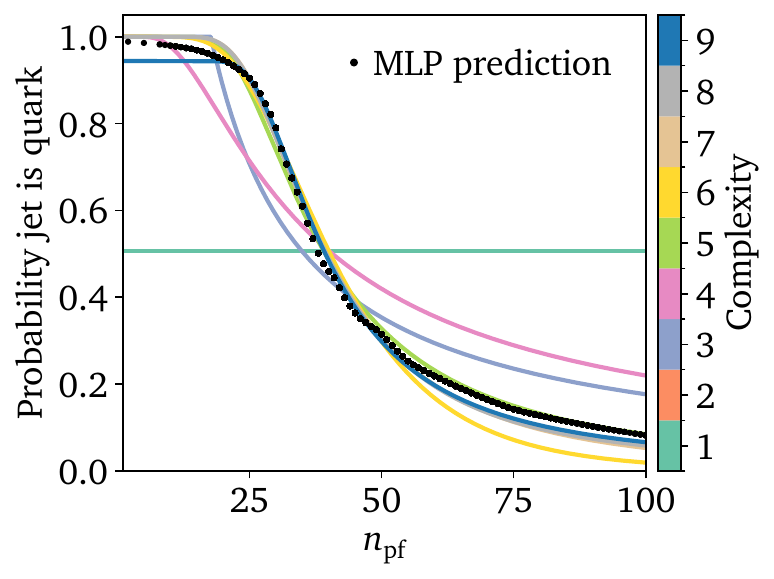}
    \includegraphics[width=0.49\linewidth]{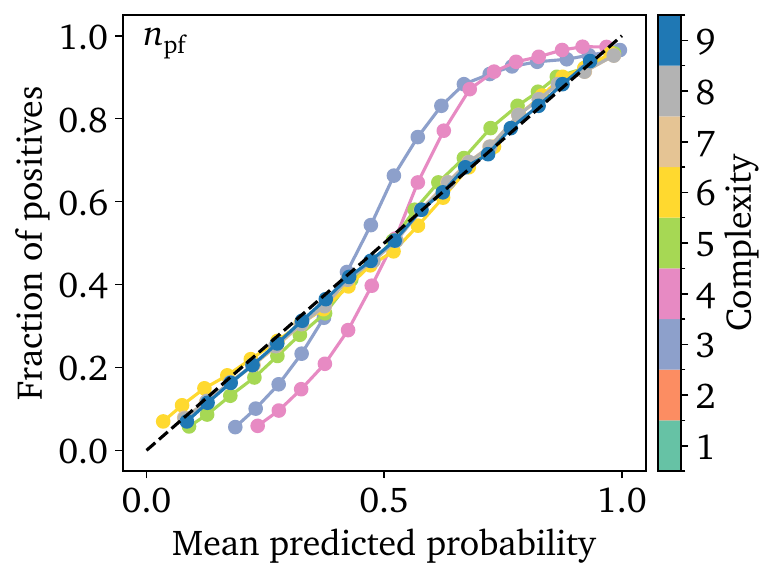}
    \caption{The left panel shows the symbolic regression curves compared to the target MLP prediction. The right panel shows the different calibration curves. A perfect calibration is shown by a diagonal (dashed black line).}
    \label{fig:npf_equations&calibrations}
\end{figure}

A subtle point arises when comparing symbolic regressions based on the AUC. Because the order of the ROC curve is invariant under monotonic transformations~\cite{Cali_2015}, formulas that differ substantially in sharpness or calibration will give identical AUC scores. This is evident in most of the one-dimensional regressions, where all formulas are monotonic transformations with identical AUCs. In Fig.~\ref{fig:npf_equations&calibrations}, we see that visually the formulas differ for $\npf$ even if the AUC remains the same. Additionally, we observe that higher complexities match the calibrated tagger prediction more closely, and we indeed require higher complexities for a good calibration. 

\begin{table}[t]
\centering
\begin{small}
\begin{tabular}{c c p{6cm}|S[table-format=1.3] S[table-format=2.2] S[table-format=2.2]}
\toprule
 observable & Complexity& Formula&  {AUC} &{$\rej_{30\%}$}& {$\Delta C$} \\
 \midrule
$\npf$ &9&$0.94 \cdot \tanh{\left(21036 \cdot\left(0.005 + \frac{1}{\npf}\right)^{3} \right)}$&0.839&52.32&1.08\\
$\sfrag$&6&$\tanh^{2} \frac{18.08}{S_{\mathrm{frag}}^{3}} $&0.828&38.97&1.80\\
$\ptd$&7&$0.92 \cdot\tanh{\left(19.49\cdot \ptd^{3} \right)}$&0.807&26.87&1.04\\
$C_{0.2}$&9&$\tanh{\left(343.13 \cdot \left(0.72 \cdot C_{0.2} - 1\right)^{18} + 0.22 \right)}$&0.793&58.41&1.54\\

$r_\lambda$&10&$\left(\left(0.59 - \tanh{\left(0.0038 \cdot r_\lambda \right)}\right)^{2}\right)^{0.5} + 0.22$&0.637&6.46&1.81\\
$\spid$ &7&$\tanh{\left(\left(S_{\mathrm{PID}} - 1.63\right)^{2} \right)} + 0.42$&0.599&6.19&1.94\\
\bottomrule
\end{tabular}
\end{small}
\caption{Best equations for each observable based on simplicity, performance and calibration. All numbers are rounded to 2 digits for readability.}
\label{tab:SR_summary}
\end{table}

\subsubsection*{Formulas for each observable}

Having validated our strategy on $\npf$, we now extend it to the full set of high-level observables
\begin{align}
\left\{ \npf, \spid, r_\lambda, \sfrag, \ptd, C_{0.2}, E_Q 
 \right\}\eqperiod
\end{align}
Following the previous sections, they span different aspects of jet substructure, including multiplicity, angular spread, fragmentation, and charge.

For each observable, we first train a classifier and then use \pysr to approximate it. Our maximum complexity is 10, to ensure the equations are interpretable. Table~\ref{tab:SR_summary} summarizes the best formulas for each observable, along with the AUC, background rejection at 30\% signal efficiency, and calibration error $\Delta C$. For each case, we select the simplest formula that achieves good performance across all three metrics. Full complexity scans for each observable are provided in the Appendix.

We can observe patterns in these equations. The fragmentation entropy  $\sfrag$, behaves similarly to $\npf$ and yields a relatively simple formula. The inverse behavior of $\npf$ resembles Casimir scaling, higher particle diversity tends to favor the gluon label, which is assigned to 0. On the other end of the spectrum, $r_\lambda$ shows limited discriminative power, and the corresponding formula is noticeably more complex. In general, more informative observables tend to produce simpler formulas, often involving only a few transformations to capture the relevant trends.

\subsection{Two-dimensional regression}

Next, we apply symbolic regression to pairs of observables. We allow for a complexity of 22 to accurately describe the 2-dimensional dependence. Rather than testing all possible combinations, we focus on pairs including $\npf$ as a standard observable.
To improve numerical stability and keep coefficient scales comparable, we multiply $\npf$ by a factor of 0.01 before regression. We deliberately avoid automated rescaling or ML-based normalization, as preserving the native physical scales of the observables supports direct interpretability of the resulting formulas.

\begin{figure}[b!]
\includegraphics[width=0.495\linewidth]{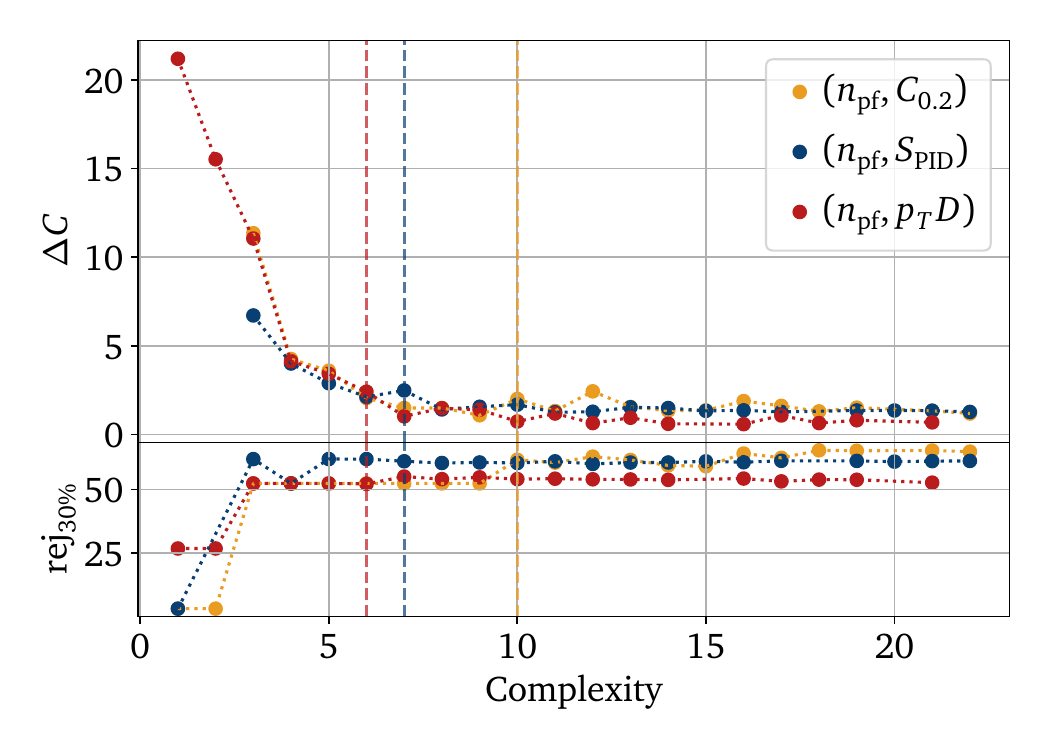}
\includegraphics[width=0.495\linewidth]{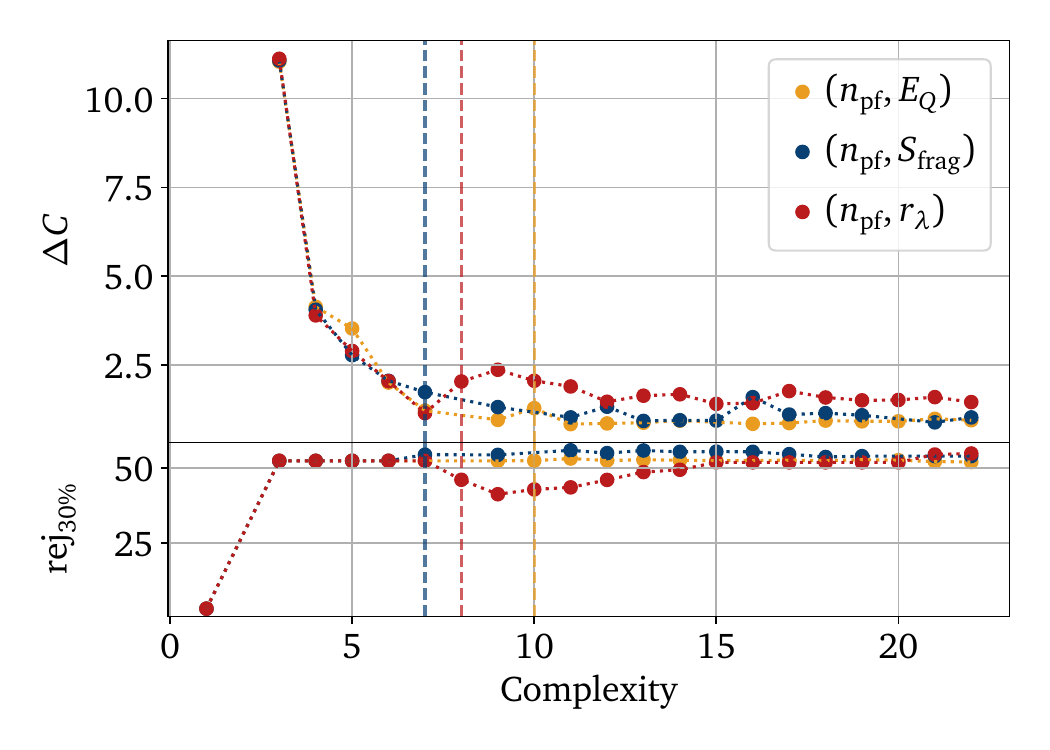}
\caption{Rejection rates at 30\% efficiency and calibration error dependent on the complexity. The vertical lines show when the equation depends for the first time on two observables. }
\label{fig:calib-2d}
\end{figure}

At low complexities, the formulas remain effectively one-dimensional. This is expected, because even a basic linear combination of two observables can already have a complexity of 7 or more. For low complexity thresholds, the second observable is too expensive to be included in the formula. In fact, up to a complexity of 4, the symbolic regression arrives at the same formulas as for $\npf$ only. The complete set of formulas, ranging from simple to complex, is provided in the Appendix. In Fig.~\ref{fig:calib-2d}, we mark by vertical lines the point at which the second observable appears in the formula. Towards higher complexity, the second observable clearly improves the performance. Table~\ref{tab:SR_summary_2D} summarizes the best formulas for each $\npf$-based pair, selected by balancing complexity, AUC, rejection rate, and calibration quality.

\begin{table}[t]
\centering
\begin{small}
\begin{tabular}{c c l|S[table-format=1.3]S[table-format=2.2]S[table-format=2.2]}
\toprule
 Obs.\,pair & Complexity& Formula&  {AUC} &{$\rej_{30\%}$}& {$\Delta C$} \\
 \midrule
$(\npf,r_\lambda)$ & 15 & $\left(1.1 - \npf\right) \cdot \tanh{\left(181.1 \cdot \left(\sqrt{\frac{1}{r_\lambda}} + \frac{0.003}{\npf^{3}}\right)^{3} \right)}$ & 0.860 & 51.81 & 1.41 
\\
$(\npf,\spid)$&9&$\tanh{\left(\frac{0.28}{n_{\mathrm{pf}}^{2} \cdot \left(n_{\mathrm{pf}} + S_{\mathrm{PID}}\right)^{2}} \right)}$&0.849&64.43&1.01\\
$(\npf, C_{0.2})$ & 18 & $\left(0.03 - \tanh^{3}\left(\frac{82.47\cdot \left(\left(C_{0.2} - 0.68\right)^{2} - 0.04\right)^{2} + 0.54}{\npf} \right)\right)^{2}$ & 0.848 & 65.27 & 1.30 
\\
$(\npf, \sfrag)$ & 13 & $0.94 \cdot \tanh^{2}{\left(\frac{0.57}{\left(\npf - 0.13\right) \cdot \left(\npf - \sfrag\right)} \right)}$ & 0.844 & 55.74 & 0.93 
\\
$(\npf, \ptd)$ & 7 & $\tanh \frac{0.26 \cdot \ptd}{\npf^{2}}$ & 0.844 & 55.01 & 1.02 
\\
$(\npf, E_Q)$ & 11 & $-0.098 \cdot E_Q + \tanh{\left(0.11 + \frac{0.03}{\npf^{3}} \right)}$ & 0.840 & 52.36 & 0.84 \\
\bottomrule
\end{tabular}
\end{small}
\caption{Symbolic regression results for pairs of observables including $\npf$. Each formula is selected based on a balance of complexity, AUC, rejection rate, and calibration.}
\label{tab:SR_summary_2D}
\end{table}

It is not always obvious which combinations of observables will yield the highest performance gain. Interestingly, $r_\lambda$ performs poorly on its own, but complements $\npf$ best and leads to an AUC of 0.860. This shows that adding information decorrelated from $\npf$ improves the AUC the most. Similarly, $C_{0.2}$ improves the rejection rate, despite its modest individual AUC. In contrast, observables like $\ptd$ and $E_Q$ produce simpler, more readable formulas that still reach competitive performance, especially in terms of calibration. This shows that combining $\npf$ with a second observable enables symbolic regression to access richer structures and yields better discrimination and equally interpretable formulas.

\subsection{Towards all-observable regression}

Finally, we turn to the question if the full ParticleNet can be approximated by a formula in terms of all seven observables from Eq.\eqref{eq:def7}. We already know that adding an additional observable typically increases the formula complexity by at least five. Covering all observables should require a complexity of 40 or more. This scales poorly in terms of computational cost and formula interpretability.
Again, to preserve interpretability we only rescale $\npf$ by a factor of 0.01 to prevent large numerical coefficients from dominating the regression.
To ensure a fair comparison, we train a new network on the same unscaled inputs. In Table~\ref{tab:SR_full_equation} we compare the performance of the learned formula to this network.

\begin{table}[b]
\centering%
\begin{tabular}{cc|ccc}
\toprule
Observables & Model & AUC & $\rej_{30\%}$ &\\
\midrule
\multirow{2}{*}{$\left(\npf, \spid, r_\lambda, \sfrag, \ptd, C_{0.2}, E_Q 
 \right)$} & MLP & 0.874&71.75\\
& \pysr & 0.874 &  68.12\\
\bottomrule
\end{tabular}
\caption{Performance of the full model using all observables. The symbolic regression formula in Eq.\eqref{eq:final_tagger_eq} has a complexity of 26.  }
\label{tab:SR_full_equation}
\end{table}

In contrast to an estimated complexity around 40, we find that a complexity of 26 yields the best trade-off between performance and interpretability. Here, the learned formula matches the performance of the network. The best-performing formula at complexity 26 is  
\begin{align}
0.89 \cdot \tanh^{3}{\left(\frac{0.008 \cdot C_{0.2}^{2}}{\ptd^{2} \cdot \left(C_{0.2} - 0.29\right)^{2}} - 0.008 \cdot \spid \cdot r_\lambda + \frac{0.56}{\npf} \right)} + 0.061\eqperiod
\label{eq:final_tagger_eq}
\end{align}
It involves five of the seven observables, $\sfrag$ and $E_Q$ are absent for the given complexity. The classifier output is a scaled 
$\tanh^3$ with a small offset, which provides a bounded non-linear mapping of a single effective score. The dominant contributions inside the activation come from three terms. The first term combines $C_{0.2}$ and 
$\ptd$ quadratically, showing sensitivity to jet fragmentation and momentum sharing. The second term is proportional to $\spid \cdot r_\lambda$, showing that correlations between particle diversity and the radiation pattern contribute to the discrimination. The final term scales as $\frac{1}{\npf}$, consistent with the strong dependence of quark–gluon separation on constituent multiplicity. Overall, the formula combines multiplicity, fragmentation, and radiation information in a compact and interpretable form.

\begin{figure}
    \centering
    \includegraphics[width=0.49\linewidth]
    {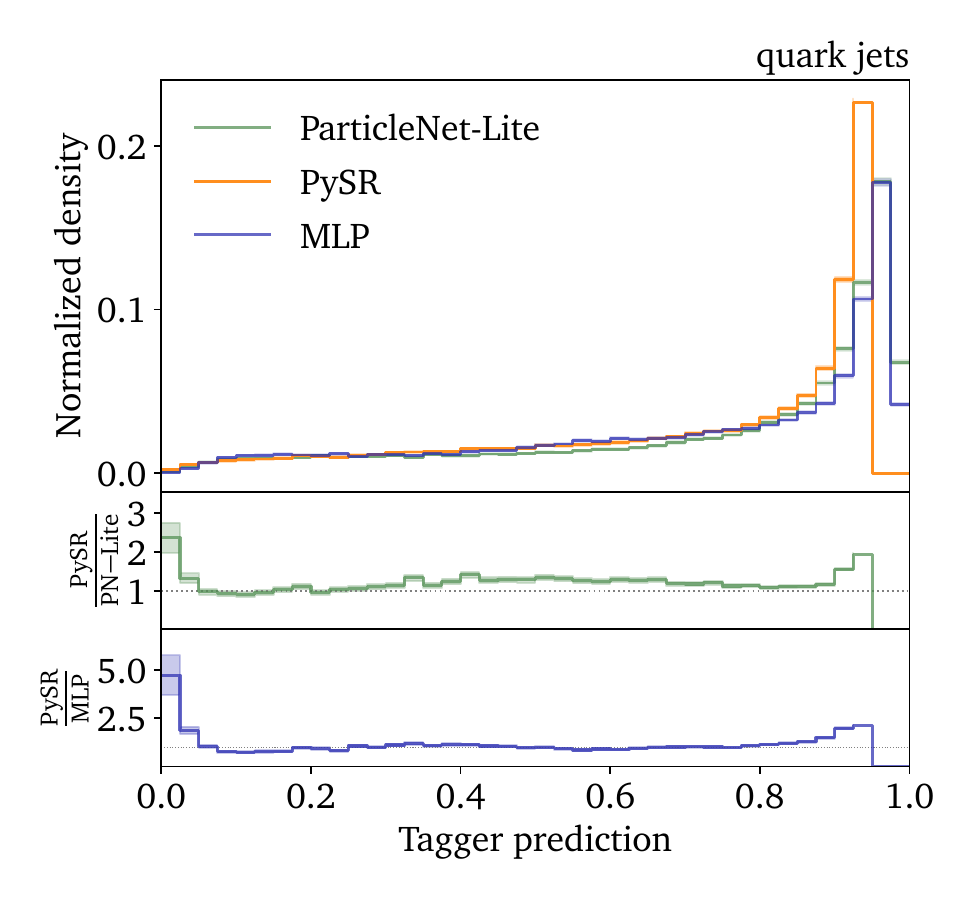}
    \includegraphics[width=0.49\linewidth]
    {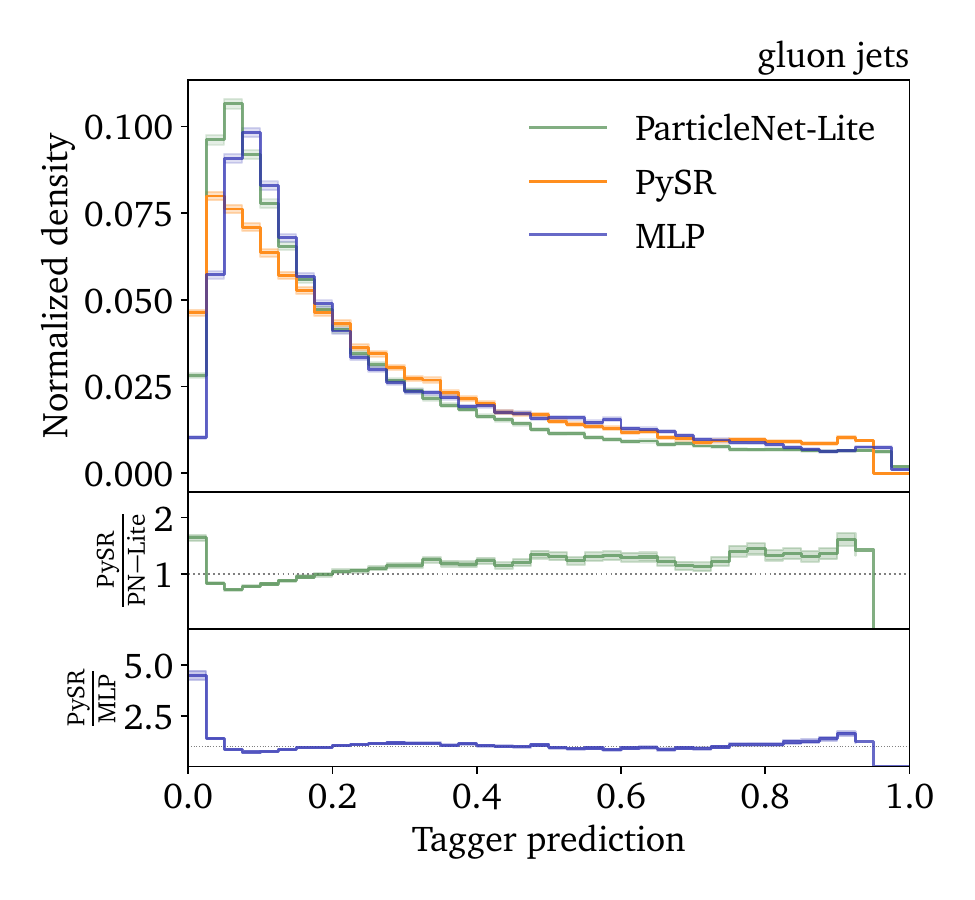}
    \caption{Score distribution plot for the MLP, \pysr and the ParticleNet-Lite predictions. The bottom two panels show the ratio between a baseline (MLP or ParticleNet Lite) and the SR formula averaged over 6 runs with error band.  }
    \label{fig:score-distribution_plot}
\end{figure}

To assess the robustness of the learned structure, we repeat the regression with \pysr five times at fixed complexity 26. We bin the classifier score into 50 equally spaced bins and compute the mean and standard deviation of the predicted score over the resulting formulas in each bin. To illustrate how the distributions change, we plot the score distributions for quark and gluon jets and compare them with the MLP and ParticleNet-Lite baselines.
Although the resulting formulas differ algebraically (see Tab.~\ref{tab:formula_robustness}), their performance is numerically indistinguishable: the AUC values agree up to negligible differences, and the bin-by-bin variation of the score is too small to be visible in Fig.~\ref{fig:score-distribution_plot}. Compared to the MLP and ParticleNet-Lite baselines, the \pysr tagger produces sharper score distributions for quark jets, with more jets assigned values close to 1 and fewer close to 0. 

\begin{table}[b!]
\begin{small}
    \centering
    \begin{tabular}{c l|c}
    \toprule
        \hspace{5pt}Trial\hspace{5pt} & Equation & AUC \\
        \midrule
        1&$0.89 \cdot \tanh^{3}{\left(\frac{0.008 \cdot C_{0.2}^{2}}{p_TD^{2} \cdot \left(C_{0.2} - 0.29\right)^{2}} - 0.008 \cdot \spid \cdot r_\lambda + \frac{0.56}{n_{\mathrm{pf}}} \right)} + 0.061$&0.874 \\
        2 &$\left(0.7 - 0.29 \cdot \tanh{\left(0.04
         \cdot p_TD \cdot r_\lambda \cdot \left(\spid + \left(C_{0.2} - 1.36\right)^{9}\right) - \frac{0.74}{n_{\mathrm{pf}}} \right)}\right)^{6}$&0.874 \\
        3&$\tanh{\left({0.27}\cdot{\left(C_{0.2} + n_{\mathrm{pf}}^{4} \cdot \left(- \spid + \frac{1.08 \cdot10^{-5} \cdot r_\lambda^{2}}{\left(- C_{0.2} - 0.03\right)^{2}} - 0.02 \cdot r_\lambda + \frac{0.37}{p_TD}\right)^{4}\right)^{-2}} \right)}$&0.874\\
        4&$1.1 - \tanh^{2}{\left(C_{0.2} + \
        r_\lambda^{2} \cdot \left(n_{\mathrm{pf}} - 0.12\right)^{2} \cdot \left(0.098\cdot p_TD - \left(C_{0.2} - 0.89\right)^{6}\right)^{2} \cdot \left(\spid - 0.14\right)^{2} \right)}$&0.874 \\
        5&$\bigg[{\tanh{\left(\frac{219 \cdot\left(\frac{1}{r_\lambda}\right)^{3/2}}{n_{\mathrm{pf}}^{3} \cdot \left(p_TD + \spid\right)^{3}} \right)} + 0.072 \cdot\tanh{\left(\frac{0.127}{0.41 - C_{0.2}} \right)}}\bigg]\cdot{(2.0 \cdot C_{0.2} + p_TD)^{-1}}$&0.874\\
\bottomrule
    \end{tabular}
    \caption{Variability of formulas at a fixed complexity of 26.}
    \label{tab:formula_robustness}
    \end{small}
\end{table}

This formula is an approximation of a tagger, meaning that it is only as robust as the tagger on which it was trained. Using a neural network, we obtain a classification prediction without explicit knowledge of how each input feature contributes. By contrast, the resulting expression provides a transparent, closed analytic form.
If we evaluate this formula on a different dataset (\eg different \pythia tunes or \herwig), its performance will reflect the robustness of the original NN classifier on which \pysr was trained. We therefore take Eq.\eqref{eq:final_tagger_eq} and evaluate it on the \herwig dataset. To provide a fair AUC comparison, we also evaluate our MLP and ParticleNet-Lite on the same dataset. The results are shown in Tab.~\ref{tab:SR_pythia_vs_herwig}.

\begin{table}[t!]
\centering%
\begin{tabular}{cc|ccc}
\toprule
observables & model & AUC & $\rej_{30\%}$ &\\
\midrule
\multirow{3}{*}{$\left(\npf, \spid, r_\lambda, \sfrag, \ptd, C_{0.2}, E_Q 
 \right)$} & MLP & 0.796&36.18\\
& \pysr Eq.\eqref{eq:final_tagger_eq}& 0.801&36.23\\
& ParticleNet-Lite & 0.804 & 37.41\\

\bottomrule
\end{tabular}
\caption{Performance of the symbolic regression for a network trained on \pythia and evaluated on \herwig, comparing symbolic regression and a MLP and ParticleNet-Lite}
\label{tab:SR_pythia_vs_herwig}
\end{table}

When evaluated on the \herwig dataset, all models exhibit noticeable performance degradation relative to the \pythia dataset. This is expected given intrinsic differences between the two generators, in particular their hadronization models: \pythia employs the Lund string model, while \herwig relies on the cluster model, leading to systematic shifts in observables such as particle multiplicity and fragmentation patterns, as also discussed in Ref.~\cite{Butter:2022xyj}. The \pysr formula performs on par with both neural-network baselines on \herwig, both in terms of AUC and background rejection. Its cross-generator behavior closely tracks that of the underlying network it approximates.

Importantly, the symbolic-regression formula provides an interpretable decomposition of the observed generator dependence. The dominant terms involve observables known to be sensitive to hadronization modeling~\cite{Butter:2022xyj}, such as constituent multiplicity and fragmentation-related features. This indicates that the performance degradation is largely associated with shifts in these inputs, rather than being entirely hidden in complex correlations of the network representation. In this sense, symbolic regression does not remove generator dependence, but offers a transparent way to illustrate how it manifests within the learned decision function.

\section{Outlook}
\label{sec:outlook}

We have shown that quark-gluon taggers trained on low-level jet constituents, despite their complexity, rely on a small set of physically meaningful features. By examining the latent representations of a trained ParticleNet-Lite, we found that much of its performance can be attributed to a few directions, closely related to (i) jet multiplicity, (ii) radial energy flow, and (iii) fragmentation. These directions are not hard-coded into the network but learned, \ie the training re-discovers the relevant physics and combines it with subtle additional structures.

Beyond confirming the established observables, our analysis suggests combinations of known observables that may be useful for future tagging studies. An example is $r_\lambda$ to isolate radial jet structure while remaining decorrelated from multiplicity. Combinations involving fragmentation entropy or charge-related observables show how ParticleNet uses information not captured by the leading substructure observables. We address the question of how to utilize particle identification, a challenge for constituent-based taggers, through its entropy $\spid$. It quantifies the diversity of particle types and is strongly correlated with one of the leading latent directions. This shows that the network leverages not just the presence of charged particles, but also the variety of particle types. Another interesting point is that the jet observables are almost exclusively linearly encoded, as suggested by our mutual information and copula study. 

Our Shapley value analysis with the \shap framework highlights both the potential and the limitations of feature attribution in jet tagging. While \shap successfully identifies physically meaningful observables, its assumption of independent inputs leads to distorted attributions in the presence of correlations. Using decorrelated input features restores consistency with physics expectations. This underscores the importance of careful input preparation when applying \shap to strongly correlated jet observables.

Finally, we employed symbolic regression using \pysr to derive simple formulas for these observables that can accurately reproduce the network output. While formulas in terms of only one observable follow the expected pattern, adding a second observable gives us valuable information about additional uncorrelated distinctive power. When allowed to use the seven leading observables, the learned formula only uses five and finds a good compromise between complexity and power. It almost matches the performance of the corresponding trained network and reflects the robustness across generators of the underlying network.

In the longer term, the compact formulas obtained through symbolic regression could be explored as fast surrogates for full ML taggers in experimental analyses. Their simple analytic form enables rapid evaluation on large-scale datasets and within environments where computational speed is a critical constraint. While they may not capture the full complexity of a network, such formulas could provide a practical compromise between performance and computational efficiency.

Our set of XAI tools provides a systematic way to understand trained precision networks without compromising training objectives or performance. By relating learned representations to well-defined physical observables, the framework moves beyond treating ML taggers as black boxes and enables a transparent interpretation of their decision-making. From a physics-analysis perspective, this allows to identify which combinations of observables drive discrimination power, offering concrete applications for Monte Carlo tuning, generator validation, and robustness studies. In particular, differences between generators or modeling assumptions can be traced back to specific physical features, providing targeted handles for systematic uncertainties and the design of more stable and interpretable tagging strategies in future analyses.


\section*{Acknowledgements}
We would like to thank Björn Malte Schäfer, Rebecca Maria Kuntz and Benedikt Schosser for the helpful discussions about mutual information. We would like to thank the Baden\hyp Württemberg\hyp Stiftung for financing through the program \textsl{Internationale Spitzenforschung}, project
\textsl{Uncertainties --- Teaching AI its Limits}
(BWST\_IF2020-010). This work is supported by the Deutsche
Forschungsgemeinschaft (DFG, German Research Foundation) under grant
396021762 -- TRR~257 \textsl{Particle Physics Phenomenology after the
Higgs Discovery}. 
The authors acknowledge support by the state of Baden-W\"urttemberg through bwHPC and the German Research Foundation (DFG) through grant no INST 39/963-1 FUGG (bwForCluster NEMO). SV is supported by the Konrad Zuse School of Excellence in Learning and Intelligent Systems (ELIZA) through the DAAD programme Konrad Zuse Schools of Excellence in Artificial Intelligence, sponsored by the Federal Ministry of Education and Research.

\clearpage
\appendix

\section{Mutual information and copula diagnostics}
\label{app:mi_copula}

We use mutual information and copulas to quantify how the latent principal
components (PCs) relate to standard jet observables. Throughout we work with
the cumulative distributions
\begin{align}
  U = F_X(X)\eqcomma \qquad V = F_Y(Y)\eqcomma
\end{align}
which are uniform on $[0,1]$. This is justified by the invariance of mutual information under strictly monotone reparameterization,
\begin{align}
  I(X;Y) = I\bigl(f(X);g(Y)\bigr)
\end{align}
for strictly increasing $f$ and $g$. We flatten the one-dimensional spectra and analyze only the dependence structure between $(U,V)$.

\subsection*{Sklar’s theorem and Gaussian copulas}

By Sklar’s theorem any joint cumulative distribution function (CDF) $F_{X,Y}$ can be written as
\begin{align}
  F_{X,Y}(x,y) = C\bigl(F_X(x),F_Y(y)\bigr)\eqcomma
\end{align}
where $F_X$ and $F_Y$ are the marginal CDFs and $C$ is a copula, \ie\ a
bivariate CDF on $[0,1]^2$ with uniform marginals. The copula $C$ contains all
information about the dependence between $X$ and $Y$, independent of the
marginal shapes. As a baseline we use the Gaussian copula. Let $(Z_1,Z_2)$ be jointly Gaussian
with zero means, unit variances, and correlation matrix
\begin{align}
  \text{P} =
  \begin{pmatrix}
    1 & \rho \\
    \rho & 1
  \end{pmatrix},
\end{align}
and define $U=\Phi(Z_1)$ and $V=\Phi(Z_2)$, with $\Phi$ the standard normal
CDF. Then $(U,V)$ has a Gaussian copula with parameter $\rho\in(-1,1)$. We say
that $(X,Y)$ has a Gaussian copula if its uniformised marginals $(U,V)$ can be
obtained in this way. For a bivariate normal $(X,Y)$ with correlation coefficient $\rho$ the mutual
information is
\begin{align}
  I(X;Y)_{\rm Gauss} = -\frac{1}{2}\log\bigl(1-\rho^2\bigr)\eqperiod
\end{align}
For a Gaussian copula the mutual information is therefore completely fixed once
$\rho$ is known; any excess mutual information beyond this expression must come
from non-Gaussian features (non-linear structure, heavy tails, asymmetries).

\subsection*{Copula families and tail dependence}

To probe non-Gaussian structure we fit standard parametric copulas to pairs
$(X,Y) = (\mathrm{PC}_i,\mathcal O_j)$, where $\mathrm{PC}_i$ are the first
five principal components of the pooled latent space and $\mathcal O_j$ denotes
a high-level jet observable such as $n_{\mathrm{PF}}$, $w_{\mathrm{PF}}$, $S_{\text{PID}}$, etc. Given a sample $\{(x_k,y_k)\}_{k=1}^n$, we first construct rank-based pseudo–observations
\begin{align}
  u_k = \frac{R_X(k)}{n+1}\eqcomma \qquad
  v_k = \frac{R_Y(k)}{n+1}\eqcomma
\end{align}
where $R_X(k)$ and $R_Y(k)$ are the empirical ranks of $x_k$ and $y_k$. This
yields approximately uniform marginals and isolates the copula. We fit the following bivariate copula families to $\{(u_k,v_k)\}$:
\begin{itemize}
  \item Gaussian copula $C_{\text{Gauss}}(u,v;\rho)$;
  \item Student-$t$ copula $C_t(u,v;\rho,\nu)$ (elliptical with power-law tails);
  \item Clayton copula (pronounced lower-tail dependence);
  \item Gumbel copula (pronounced upper-tail dependence);
  \item Frank copula (no asymptotic tail dependence, flexible in the bulk).
\end{itemize}
Parameters $\theta$ are obtained by maximising the copula log-likelihood
\begin{align}
  \ell(\theta) = \sum_{k=1}^n \log c_\theta(u_k,v_k)\eqcomma
\end{align}
where $c_\theta$ is the copula density. We initialise from Kendall’s rank
correlation,
\begin{align}
  \tau = \mathrm{Kendall}(U,V)\eqcomma
\end{align}
which for elliptical copulas fixes the linear correlation $\rho$; for the
Gaussian copula
\begin{align}
  \rho = \sin\left(\frac{\pi}{2}\tau\right)\eqperiod
\end{align}
Model selection is based on the Akaike information criterion
\begin{align}
  \mathrm{AIC} = -2\ell + 2k\eqcomma
\end{align}
with the Bayesian information criterion
\begin{align}
  \mathrm{BIC} = -2\ell + k\log n
\end{align}
used as a cross-check, where $k$ is the number of free parameters. The copula
family with the lowest AIC is taken as preferred. To quantify extreme correlations we extract the upper and lower tail-dependence
coefficients
\begin{align}
  \lambda_U
    &= \lim_{q\to 1^-} \mathbb P\big[V > q \,\big|\, U > q\big]\eqcomma\\
  \lambda_L
    &= \lim_{q\to 0^+} \mathbb P\big[V \le q \,\big|\, U \le q\big]\eqcomma
\end{align}
whenever the limits exist. Nonzero $\lambda_{U,L}$ means that very large (or
very small) values of the two variables occur together more often than for
weakly correlated Gaussian variables. For the bivariate Student-$t$ copula with correlation $\rho$ and $\nu$ degrees
of freedom one has
\begin{align}
  \lambda_U = \lambda_L
    = 2\, T_{\nu+1}\!\left(
        -\sqrt{\frac{(\nu+1)(1-\rho)}{1+\rho}}
      \right),
\end{align}
where $T_{\nu+1}$ is the CDF of a univariate $t$-distribution with $\nu+1$
degrees of freedom. In contrast, for Gaussian and Frank copulas one finds
$\lambda_L=\lambda_U=0$.

\subsection*{Fit setup and observables}

We apply this procedure to $n = 10^5$ jets for all pairs
$(\mathrm{PC}_i,\mathcal O_j)$ with $i=1,\dots,5$ and
\begin{align}
  \{\mathcal O_j\}
    = \{n_{\mathrm{PF}}, S_{\mathrm{PID}}, w_{\mathrm{PF}}, r_\lambda,
       S_{\mathrm{frag}}^\perp, S_{\mathrm{frag}},
       \log p_T^D, C_{0.2}, C_{0.2}^\perp, E_Q\}.
\end{align}
For each pair we record:
\begin{itemize}
  \item the best-fit copula family (AIC and BIC);
  \item Kendall’s $\tau$ and Spearman’s $\rho_S$;
  \item the copula parameters (\eg $\rho$, $\nu$, $\theta$);
  \item the implied tail coefficients $(\lambda_L,\lambda_U)$;
  \item and a non-parametric estimate of $I(X;Y)$ from $(U,V)$, compared to the
        Gaussian baseline $I_{\rm Gauss}(X;Y)$.
\end{itemize}
In the main text we use the mutual information only as a qualitative check:
agreement with $I_{\rm Gauss}$ means that the dependence is effectively
Gaussian at the copula level.

\subsection*{Results for latent PCs and observables}

For the leading component $\mathrm{PC}_1$ we find:

\paragraph{Multiplicity  axis.}
The strongest correlation is with multiplicity  For $(\mathrm{PC}_1,n_{\mathrm{PF}})$
the best copula is Gaussian, with
\begin{align}
  \tau(\mathrm{PC}_1,n_{\mathrm{PF}}) \simeq 0.82\eqcomma
\end{align}
and $\lambda_L = \lambda_U = 0$. The mutual information is well described by
the Gaussian prediction. This shows that $\mathrm{PC}_1$ is essentially a
monotonic multiplicity axis with an almost purely Gaussian dependence
structure.

\paragraph{Fragmentation / PID observables.}
Observables such as $S_{\mathrm{PID}}$, $S_{\mathrm{frag}}$, and $C_{0.2}$ are
also strongly correlated with $\mathrm{PC}_1$:
\begin{align}
  \tau(\mathrm{PC}_1,S_{\mathrm{PID}}) \simeq 0.76,\quad
  \tau(\mathrm{PC}_1,S) \simeq 0.70,\quad
  \tau(\mathrm{PC}_1,C_{0.2}) \simeq 0.69\eqperiod
\end{align}
The preferred copulas are Frank or Gaussian with
$\lambda_L = \lambda_U = 0$, \ie\ smooth, essentially Gaussian-like
dependence without asymptotic tail enhancement.

\paragraph{Width and heavy tails.}
The jet width $w_{\mathrm{PF}}$ is special. For $(\mathrm{PC}_1,w_{\mathrm{PF}})$
the preferred family is a Student-$t$ copula with
\begin{align}
  \tau(\mathrm{PC}_1,w_{\mathrm{PF}}) \simeq 0.63,\qquad
  \lambda_L \simeq \lambda_U \simeq 0.39,\qquad
  \nu \simeq 8\eqperiod
\end{align}
The empirical mutual information significantly exceeds the Gaussian baseline,
indicating a strong monotonic dependence with sizeable, symmetric tail
dependence: very extreme values of $\mathrm{PC}_1$ and $w_{\mathrm{PF}}$
co-occur much more often than in the Gaussian case. At first sight this looks
like a genuine “width tail’’ direction in the latent space.

\paragraph{Small lower-tail effects.}
A few weakly correlated observables show Clayton behaviour with small but
nonzero lower-tail coefficients, e.g.
\begin{align}
  \tau(\mathrm{PC}_1,S_{\mathrm{frag}}^\perp) \simeq 0.02,\quad
  \tau(\mathrm{PC}_1,C_{0.2}^\perp) \simeq 0.01,\quad
  \tau(\mathrm{PC}_1,\text{neg.\ energy frac}) \simeq 0.08,
\end{align}
with $\lambda_L \sim 0.03$–$0.07$ and $\lambda_U \simeq 0$. These encode a mild
preference for extreme low-end configurations (\eg unusually large negative
energy contributions) but do not affect the bulk.

For the subleading components $\mathrm{PC}_{2-5}$ the picture is much milder:
\begin{itemize}
  \item A few pairs show moderate correlations, notably
    $(\mathrm{PC}_2,\Lambda)$ with $\tau\simeq 0.58$ (Gaussian copula, no tails)
    and $(\mathrm{PC}_3,S_{\mathrm{frag}}^\perp)$ and
    $(\mathrm{PC}_3,C_{0.2}^\perp)$ with $\tau\simeq 0.48$–$0.47$
    (Gaussian/Frank, no tails). These are again essentially Gaussian at the
    copula level.
  \item Most other pairs involving $\mathrm{PC}_{2-5}$ have
    $|\tau|\lesssim 0.2$. The best-fit copulas are predominantly Gaussian or
    Frank with $\lambda_{L,U}=0$. In the few cases where AIC prefers Student-$t$
    or Clayton families, the tail coefficients are at the level
    $|\lambda_{L,U}|\lesssim \mathcal O(10^{-2})$.
\end{itemize}
Overall, $\mathrm{PC}_1$ is cleanly identified as a monotonic multiplicity axis with mostly Gaussian dependence on standard observables. Higher PCs carry weaker and more specialised structures, largely without
significant tail dependence.

\subsection*{Decorrelated width versus multiplicity}

The heavy tails in $(\mathrm{PC}_1,w_{\mathrm{PF}})$ can be largely explained by
the trivial statement that wider jets tend to have larger multiplicity. To
isolate genuine width information we use a decorrelated width observable
$w_{\mathrm{PF}}^\perp$, defined in the main text as a linear combination
\begin{align}
  w_{\mathrm{PF}}^\perp = \alpha\, n_{\mathrm{PF}} - w_{\mathrm{PF}}\eqcomma
\end{align}
with $\alpha$ chosen such that $w_{\mathrm{PF}}^\perp$ is (approximately)
uncorrelated with $n_{\mathrm{PF}}$. This removes the leading multiplicity
dependence while keeping sensitivity to the radial energy profile.

Repeating the copula analysis with $w_{\mathrm{PF}}^\perp$ (or, equivalently, a
residualised width obtained from regressing $w_{\mathrm{PF}}$ on $n_{\mathrm{PF}}$)
we find that the best-fit copula is still Student-$t$, but the dependence on
$\mathrm{PC}_1$ is now very weak:
\begin{align}
  \tau(\mathrm{PC}_1,w_{\mathrm{PF}}^\perp)
    \simeq 0.06,\qquad
  \lambda_L \simeq \lambda_U \simeq 0.02\eqperiod
\end{align}
The mutual information drops by about an order of magnitude and becomes close
to the Gaussian prediction based on the same $\rho$.

We conclude that essentially all of the strong, heavy-tailed correlation
between $\mathrm{PC}_1$ and jet width is mediated by multiplicity. Once
$w_{\mathrm{PF}}$ is decorrelated from $n_{\mathrm{PF}}$ (via
$w_{\mathrm{PF}}^\perp$), the remaining dependence of $\mathrm{PC}_1$ on width is
small and only mildly non-Gaussian. In other words, the apparent “width tail’’
of $\mathrm{PC}_1$ is really just the multiplicity tail seen through $w_{\mathrm{PF}}$.

\clearpage
\section{Supplementary tables}

\begin{table}[ht]
\centering
\begin{small}
\begin{small}
\begin{tabular}{l|l|l|l|l|l}
\toprule
PCA Combination & AUC & $\rej_{30\%}$ & PCA Combination & AUC & $\rej_{30\%}$ \\
\midrule
$\text{PC}_1$, $\text{PC}_2$, $\text{PC}_3$, $\text{PC}_4$, $\text{PC}_5$ & 0.901 & 93.33 & $\text{PC}_2$, $\text{PC}_3$, $\text{PC}_4$, $\text{PC}_5$ & 0.774 & 17.36 \\
$\text{PC}_1$, $\text{PC}_2$, $\text{PC}_4$, $\text{PC}_5$ & 0.898 & 78.60 & $\text{PC}_2$, $\text{PC}_4$, $\text{PC}_5$ & 0.733 & 10.67 \\
$\text{PC}_1$, $\text{PC}_2$, $\text{PC}_3$, $\text{PC}_5$ & 0.898 & 87.84 & $\text{PC}_3$, $\text{PC}_4$, $\text{PC}_5$ & 0.728 & 12.08 \\
$\text{PC}_1$, $\text{PC}_2$, $\text{PC}_5$ & 0.896 & 86.15 & $\text{PC}_2$, $\text{PC}_3$, $\text{PC}_4$ & 0.727 & 9.80 \\
$\text{PC}_1$, $\text{PC}_2$, $\text{PC}_3$, $\text{PC}_4$ & 0.893 & 78.60 & $\text{PC}_2$, $\text{PC}_3$, $\text{PC}_5$ & 0.718 & 10.87 \\
$\text{PC}_1$, $\text{PC}_3$, $\text{PC}_4$, $\text{PC}_5$ & 0.893 & 81.45 & $\text{PC}_2$, $\text{PC}_4$ & 0.686 & 7.49 \\
$\text{PC}_1$, $\text{PC}_2$, $\text{PC}_3$ & 0.891 & 74.67 & $\text{PC}_2$, $\text{PC}_5$ & 0.681 & 7.13 \\
$\text{PC}_1$, $\text{PC}_4$, $\text{PC}_5$ & 0.891 & 73.44 & $\text{PC}_3$, $\text{PC}_4$ & 0.676 & 7.79 \\
$\text{PC}_1$, $\text{PC}_2$, $\text{PC}_4$ & 0.890 & 75.93 & $\text{PC}_2$, $\text{PC}_3$ & 0.668 & 7.31 \\
$\text{PC}_1$, $\text{PC}_3$, $\text{PC}_5$ & 0.889 & 78.60 & $\text{PC}_4$, $\text{PC}_5$ & 0.665 & 7.70 \\
$\text{PC}_1$, $\text{PC}_2$ & 0.888 & 80.00 & $\text{PC}_2$ & 0.634 & 5.57 \\
$\text{PC}_1$, $\text{PC}_5$ & 0.887 & 72.26 & $\text{PC}_3$, $\text{PC}_5$ & 0.627 & 7.32 \\
$\text{PC}_1$, $\text{PC}_3$, $\text{PC}_4$ & 0.886 & 75.93 & $\text{PC}_4$ & 0.611 & 5.45 \\
$\text{PC}_1$, $\text{PC}_3$ & 0.883 & 74.67 & $\text{PC}_3$ & 0.575 & 5.42 \\
$\text{PC}_1$, $\text{PC}_4$ & 0.882 & 73.44 & $\text{PC}_5$ & 0.549 & 4.18 \\
$\text{PC}_1$ & 0.879 & 74.67 & & & \\
\bottomrule
\end{tabular}
\end{small}
\caption{Feedforward classifier trained on different PC combinations using the \pythia dataset, showing AUC and 30\% quark rejection rate.}
\label{tab:pca_combination_auc_rej30_6col}
\end{small}
\label{tab:pca_sorted_split}
\end{table}

\begin{table}[ht]
\centering
\begin{small}
\begin{tabular}{l|l|l|l|l|l}
\toprule
PCA Combination & AUC & $\rej_{30\%}$ & PCA Combination & AUC & $\rej_{30\%}$ \\
\midrule$\text{PC}_1$, $\text{PC}_2$, $\text{PC}_3$, $\text{PC}_4$, $\text{PC}_5$ & 0.831 & 47.19 & 
 $\text{PC}_2$, $\text{PC}_3$, $\text{PC}_4$, $\text{PC}_5$ & 0.715 & 11.49 \\
$\text{PC}_1$, $\text{PC}_2$, $\text{PC}_3$, $\text{PC}_5$ & 0.831 & 47.16 & $\text{PC}_2$, $\text{PC}_3$, $\text{PC}_4$ & 0.692 & 9.89 \\
$\text{PC}_1$, $\text{PC}_2$, $\text{PC}_3$, $\text{PC}_4$ & 0.831 & 46.19 & $\text{PC}_2$, $\text{PC}_3$, $\text{PC}_5$ & 0.692 & 9.91 \\
$\text{PC}_1$, $\text{PC}_2$, $\text{PC}_4$, $\text{PC}_5$ & 0.831 & 44.36 & $\text{PC}_2$, $\text{PC}_3$ & 0.667 & 8.89 \\
$\text{PC}_1$, $\text{PC}_2$, $\text{PC}_3$ & 0.831 & 45.71 & $\text{PC}_2$, $\text{PC}_4$, $\text{PC}_5$ & 0.654 & 7.87 \\
$\text{PC}_1$, $\text{PC}_2$, $\text{PC}_4$ & 0.830 & 44.36 & $\text{PC}_2$, $\text{PC}_4$ & 0.646 & 7.47 \\
$\text{PC}_1$, $\text{PC}_2$, $\text{PC}_5$ & 0.830 & 45.71 & $\text{PC}_3$, $\text{PC}_4$, $\text{PC}_5$ & 0.637 & 6.54 \\
$\text{PC}_1$, $\text{PC}_2$ & 0.830 & 45.71 & $\text{PC}_3$, $\text{PC}_5$ & 0.629 & 6.25 \\
$\text{PC}_1$, $\text{PC}_3$, $\text{PC}_4$, $\text{PC}_5$ & 0.814 & 43.92 & $\text{PC}_3$, $\text{PC}_4$ & 0.620 & 5.93 \\
$\text{PC}_1$, $\text{PC}_3$, $\text{PC}_4$ & 0.814 & 45.71 & $\text{PC}_2$, $\text{PC}_5$ & 0.615 & 6.35 \\
$\text{PC}_1$, $\text{PC}_3$, $\text{PC}_5$ & 0.814 & 44.80 & $\text{PC}_3$ & 0.611 & 5.43 \\
$\text{PC}_1$, $\text{PC}_3$ & 0.813 & 44.36 & $\text{PC}_2$ & 0.609 & 6.22 \\
$\text{PC}_1$, $\text{PC}_4$, $\text{PC}_5$ & 0.812 & 49.23 & $\text{PC}_4$, $\text{PC}_5$ & 0.553 & 4.44 \\
$\text{PC}_1$, $\text{PC}_5$ & 0.811 & 47.66 & $\text{PC}_4$ & 0.537 & 3.76 \\
$\text{PC}_1$, $\text{PC}_4$ & 0.811 & 48.17 & $\text{PC}_5$ & 0.530 & 3.90 \\
$\text{PC}_1$ & 0.811 & 47.66 & & & \\
\bottomrule
\end{tabular}
\end{small}

\caption{Feedforward classifier trained on different PC combinations using the \herwig dataset, showing AUC and 30\% quark rejection rate }
\label{tab:pca_test_auc_herwig}
\end{table}

\begin{table}[ht]
\centering
\begin{small}
\begin{tabular}{l|l|l|l|l|l}
\toprule
PCA Combination & AUC & $\rej_{30\%}$ & PCA Combination & AUC & $\rej_{30\%}$ \\
\midrule
$\text{PC}_1$, $\text{PC}_2$, $\text{PC}_3$, $\text{PC}_4$, $\text{PC}_5$ & 0.829 & 41.87 & $\text{PC}_3$, $\text{PC}_4$ & 0.805 & 43.08 \\
$\text{PC}_1$, $\text{PC}_2$, $\text{PC}_3$, $\text{PC}_4$ & 0.828 & 43.08 & $\text{PC}_1$, $\text{PC}_2$, $\text{PC}_4$ & 0.805 & 23.33 \\
$\text{PC}_1$, $\text{PC}_2$, $\text{PC}_3$, $\text{PC}_5$ & 0.828 & 38.62 & $\text{PC}_1$, $\text{PC}_2$ & 0.803 & 22.97 \\
$\text{PC}_2$, $\text{PC}_3$, $\text{PC}_4$, $\text{PC}_5$ & 0.828 & 43.08 & $\text{PC}_2$, $\text{PC}_4$ & 0.803 & 22.40 \\
$\text{PC}_2$, $\text{PC}_3$, $\text{PC}_5$ & 0.828 & 40.00 & $\text{PC}_2$ & 0.801 & 21.85 \\
$\text{PC}_1$, $\text{PC}_2$, $\text{PC}_3$ & 0.827 & 39.65 & $\text{PC}_3$, $\text{PC}_5$ & 0.801 & 34.20 \\
$\text{PC}_2$, $\text{PC}_3$, $\text{PC}_4$ & 0.827 & 41.87 & $\text{PC}_1$, $\text{PC}_3$ & 0.783 & 32.46 \\
$\text{PC}_2$, $\text{PC}_3$ & 0.826 & 39.30 & $\text{PC}_3$ & 0.780 & 32.70 \\
$\text{PC}_1$, $\text{PC}_3$, $\text{PC}_4$, $\text{PC}_5$ & 0.816 & 42.26 & $\text{PC}_1$, $\text{PC}_4$, $\text{PC}_5$ & 0.767 & 22.51 \\
$\text{PC}_3$, $\text{PC}_4$, $\text{PC}_5$ & 0.816 & 42.67 & $\text{PC}_4$, $\text{PC}_5$ & 0.764 & 20.55 \\
$\text{PC}_1$, $\text{PC}_2$, $\text{PC}_4$, $\text{PC}_5$ & 0.816 & 33.43 & $\text{PC}_1$, $\text{PC}_4$ & 0.759 & 17.78 \\
$\text{PC}_1$, $\text{PC}_3$, $\text{PC}_4$ & 0.813 & 42.67 & $\text{PC}_4$ & 0.756 & 17.85 \\
$\text{PC}_1$, $\text{PC}_2$, $\text{PC}_5$ & 0.811 & 27.48 & $\text{PC}_1$, $\text{PC}_5$ & 0.679 & 9.45 \\
$\text{PC}_2$, $\text{PC}_4$, $\text{PC}_5$ & 0.810 & 28.00 & $\text{PC}_1$ & 0.641 & 7.09 \\
$\text{PC}_2$, $\text{PC}_5$ & 0.809 & 24.75 & $\text{PC}_5$ & 0.563 & 4.93 \\
$\text{PC}_1$, $\text{PC}_3$, $\text{PC}_5$ & 0.807 & 36.72 & & & \\
\bottomrule
\end{tabular}
\end{small}
\caption{Feedforward classifier trained on different PC combinations using the \pythia PCA directions applied to the \herwig dataset, showing AUC and 30\% quark rejection rate }
\label{tab:pca_test_auc_herwig_pythia_mix}
\end{table}


\begin{table}
\centering
\begin{small}
\begin{tabular}{c|p{8cm}|S[table-format=1.5]|S[table-format=1.3]|S[table-format=2.2]|S[table-format=2.2]}
\toprule
Complexity& Formula& {Loss}& {AUC} &{$\rej_{30\%}$}& {$\Delta C$} \\
 \midrule
1&0.5&0.083&0.500&3.33&{-} \\
4&$\tanh{\left(\frac{0.78}{S_{\mathrm{PID}}} \right)}$&0.0043&0.594&6.19& 12.75\\
5&$1.3 - \tanh^{2}{\left(S_{\mathrm{PID}} \right)}$&0.0036&0.594&6.19&7.11\\
6&$\left(S_{\mathrm{PID}} - 1.6\right)^{2} + 0.43$&0.00089&0.599&6.20&2.30\\
7&$\tanh{\left(\left(S_{\mathrm{PID}} - 1.63\right)^{2} \right)} + 0.42$&0.00056&0.599&6.19&1.94\\
8&$\tanh{\left(\left(\left(S_{\mathrm{PID}} - 1.61\right)^{2} + 0.67\right)^{2} \right)}$&0.00009&0.599&6.19&1.64 \\
10&$0.99 \cdot \tanh{\left(\left(\left(S_{\mathrm{PID}} - 1.61\right)^{2} + 0.68\right)^{2} \right)}$&0.00009&0.599&6.19&2.00\\
     \bottomrule
\end{tabular}
\end{small}
\caption{1D symbolic regression tables for $\spid$}
\end{table}

\begin{table}
\centering
\begin{small}
\begin{tabular}{c|p{8cm}|S[table-format=1.5]|S[table-format=1.3]|S[table-format=2.2]|S[table-format=2.2]}
    \toprule
     Complexity& Formula& {Loss}& {AUC} &{$\rej_{30\%}$}& {$\Delta C$} \\
     \midrule
1&0.5&0.014&0.5&3.33& {-}\\
4&$4.3 \cdot \left(\frac{1}{r_\lambda}\right)^{0.5}$&0.003&0.637&6.46&7.39\\
5&$0.78 - 0.0032 \cdot r_\lambda$&0.0003&0.637&6.46&4.52\\
6&$0.85 \cdot\left(1.0 - 0.0019 \cdot r_\lambda\right)^{3}$&0.0001&0.637&6.46&3.00\\
7&$\left(0.92 - \tanh{\left(0.002\cdot r_\lambda \right)}\right)^{2}$&0.00011&0.637&6.46&2.80\\
8&$\tanh{\left(\frac{114.82}{r_\lambda + 36.44} \right)} - 0.23$&0.00009&0.637&6.46&2.31\\
9&$0.78 \cdot\left(\left(1.0 - 0.0051 \cdot r_\lambda\right)^{2} + 0.099\right)^{0.5}$&0.00005&0.637&6.46&2.03\\
10&$\left(\left(0.59 - \tanh{\left(0.0038 \cdot r_\lambda \right)}\right)^{2}\right)^{0.5} + 0.22$&0.00005&0.637&6.46&1.81\\
     \bottomrule
\end{tabular}
\end{small}
\caption{1D symbolic regression tables for $r_\lambda$}
\end{table}

\begin{table}
\centering
\begin{small}
\begin{tabular}{c|p{8cm}|S[table-format=1.5]|S[table-format=1.3]|S[table-format=2.2]|S[table-format=2.2]}
    \toprule
     Complexity& Formula& {Loss}& {AUC} &{$\rej_{30\%}$}& {$\Delta C$} \\
     \midrule
1&$\ptd$&0.058&0.807&26.87&21.19\\
2&$\ptd^{0.5}$&0.038&0.807&26.87&15.52\\
3&$1.6 \cdot \ptd$&0.015&0.807&26.87&10.78\\
5&$\tanh{\left(17.17\cdot \ptd^{3} \right)}$&0.0009&0.807&26.87&2.15\\
6&$\tanh^{9}{\left(5.25\cdot \ptd \right)}$&0.0005&0.807&26.87&1.81\\
7&$0.92 \cdot\tanh{\left(19.49\cdot \ptd^{3} \right)}$&0.00014&0.807&26.87&1.04\\
9&$0.92 \cdot\tanh{\left(22.1 \cdot \left(\ptd - 0.01\right)^{3} \right)}$&0.00011&0.807&26.87&0.87\\
10&$\tanh^{2}{\left(21.33 \cdot \ptd^{3} + 0.35 \right)} - 0.087$&0.00010&0.807&26.87&0.80\\
\bottomrule
\end{tabular}
\end{small}
\caption{1D symbolic regression tables for $\ptd$}
\end{table}

\begin{table}
\centering
\begin{small}
\begin{tabular}{c|p{8cm}|S[table-format=1.5]|S[table-format=1.3]|S[table-format=2.2]|S[table-format=2.2]}
\toprule
 Complexity& Formula& {Loss}& {AUC} &{$\rej_{30\%}$}& {$\Delta C$} \\
 \midrule
1&$0.5$&0.076&0.5&3.33& {--}\\
3&$0.97 - C_{0.2}$&0.033&0.793&58.41&15.35\\
4&$2.0 \cdot \left(1.0 - 0.79 \cdot C_{0.2}\right)^{3}$&0.012&0.793&58.41&10.26\\
5&$2.8 \cdot \left(0.77 \cdot C_{0.2} - 1.0\right)^{4}$&0.010&0.793&58.41&8.09\\
6&$\tanh{\left(19.66\cdot \left(1 - 0.72\cdot C_{0.2}\right)^{9} \right)}$&0.0042&0.793&58.41&5.78\\
7&$\tanh{\left(\frac{0.026}{\left(0.23 - C_{0.2}\right)^{2}} \right)}$&0.0014&0.793&58.41&4.28\\
9&$\tanh{\left(343.13 \cdot \left(0.72 \cdot C_{0.2} - 1\right)^{18} + 0.22 \right)}$&0.00040&0.793&58.41&1.54\\
10&$\tanh{\left(88.98 \cdot\sqrt{\left(1 - 0.72 \cdot C_{0.2}\right)^{27} + 6.72 \cdot 10^{-6}} \right)}$&0.00038&0.793&58.41& 1.72\\
\bottomrule
\end{tabular}
\end{small}
\caption{1D symbolic regression tables for $C_{0.2}$}
\end{table}
\begin{table}
\centering
\begin{small}
\begin{tabular}{c|p{8cm}|S[table-format=1.5]|S[table-format=1.3]|S[table-format=2.2]|S[table-format=2.2]}
\toprule
 Complexity& Formula& {Loss}& {AUC} &{$\rej_{30\%}$}& {$\Delta C$} \\
 \midrule
1&$0.49$&0.01183&0.5&3.33& {-}\\
5&$0.43 \cdot \left(\frac{1}{E_Q}\right)^{0.25}$&0.01108&0.483&4.15&10.45\\
6&$E_Q^{3} - E_Q + 0.82$&0.00206&0.617&7.37&8.39\\
7&$E_Q^{4} - E_Q + 0.9$&0.00060&0.621&7.66&6.01\\
8&$\tanh{\left(E_Q^{3} + 1.75 \cdot \left(1 - 0.83 \cdot E_Q\right)^{3} \right)}$&0.00046&0.621&7.66&6.01\\
9&$1.2 \cdot E_Q^{4} - 1.2 \cdot E_Q + 0.99$&0.00017&0.62&7.66&2.17\\
10&$\left(E_Q - 0.031\right)^{3} + \tanh{\left(1.76 \cdot \left(1 - 0.84\cdot E_Q\right)^{3} \right)}$&0.00005&0.62&7.65&2.57\\

\bottomrule
\end{tabular}
\end{small}
\caption{1D symbolic regression tables for $E_Q$}
\end{table}
\begin{table}
\centering
\begin{small}
\begin{tabular}{c|p{8cm}|S[table-format=1.5]|S[table-format=1.3]|S[table-format=2.2]|S[table-format=2.2]}
\toprule
 Complexity& Formula& {Loss}& {AUC} &{$\rej_{30\%}$}& {$\Delta C$} \\
 \midrule
1&$0.5$&0.081&0.5&3.33&7.34\\
4&$- \tanh{\left(\sfrag - 3.5 \right)}$&0.017&0.928&38.97&3.59\\
5&$- \tanh^{3}{\left(\sfrag - 3.9\right)}$&0.0027&0.928&38.97&3.59\\
6&$\tanh^{2}{\left(\frac{18.08}{\sfrag^{3}} \right)}$&0.0005&0.928&38.97&1.80\\
7&$\tanh{\left(19.94 \cdot \left(0.2 - \frac{1}{\sfrag}\right)^{2} \right)}$&0.0005&0.928&38.97&1.50\\
8&$\left(\tanh{\left(\frac{19.38}{\sfrag^{3}} \right)} - 0.026\right)^{2}$&0.00015&0.928&38.97&3.02\\
10&$\tanh{\left(\frac{5.18}{1.71\cdot \left(0.87 \cdot \sfrag - 1\right)^{4} + 2.8} \right)}$&0.00009&0.928&38.88&0.93\\
\bottomrule
\end{tabular}
\end{small}
\caption{1D symbolic regression tables for $\sfrag$}
\end{table}

\begin{table}
\centering

\begin{small}
\begin{scriptsize}
\begin{tabular}{c|p{9cm}|S[table-format=1.5]|S[table-format=1.3]|S[table-format=2.2]|S[table-format=2.2]}
\toprule
 Complexity& Formula& {Loss}& {AUC} &{$\rej_{30\%}$}& {$\Delta C$} \\
 \midrule
 1&0.51&0.08625&0.50&3.33& {-}\\
3&$0.93 - n_{\mathrm{pf}}$&0.00832&0.840&52.32&11.04\\
4&$\left(1.2 - n_{\mathrm{pf}}\right)^{3}$&0.00656&0.840&52.32&4.16\\
5&$\tanh{\left(\frac{0.08}{n_{\mathrm{pf}}^{2}} \right)}$&0.00330&0.840&52.32&3.45\\
6&$\tanh^{6}{\left(\frac{0.6}{n_{\mathrm{pf}}} \right)}$&0.00279&0.840&52.32&2.00\\
7&$\left(- n_{\mathrm{pf}}^{2} \cdot S_{\mathrm{PID}} + 1.0\right)^{3}$&0.00047&0.848&62.94&4.11\\
8&$\tanh{\left(0.06\cdot \left(\frac{1}{S_{\mathrm{PID}}}\right)^{3/2}n_{\mathrm{pf}}^{-3}\right)}$&0.00037&0.848&62.94&2.42\\
9&$\tanh{\left(\frac{0.28}{n_{\mathrm{pf}}^{2} \cdot \left(n_{\mathrm{pf}} + S_{\mathrm{PID}}\right)^{2}} \right)}$&0.00035&0.849&64.43&1.01\\
10&$\tanh{\left(\frac{0.1 \cdot \left(\frac{1}{S_{\mathrm{PID}}}\right)^{3/2}}{\left(n_{\mathrm{pf}} + 0.09\right)^{3}} \right)}$&0.00020&0.849&64.28&0.96\\
11&$0.96 \times \tanh{\left(\frac{0.3}{n_{\mathrm{pf}}^{2} \times \left(n_{\mathrm{pf}} + S_{\mathrm{PID}}\right)^{2}} \right)}$&0.00019&0.849&64.43&1.11\\
12&$0.96 \cdot \tanh{\left(\frac{0.095 \cdot\left(\frac{1}{S_{\mathrm{PID}}}\right)^{3/2}}{\left(n_{\mathrm{pf}} + 0.06\right)^{3}} \right)}$&0.00012&0.849&63.84&0.73\\
13&$0.96 \cdot \tanh^{1.5}{\left(\frac{0.67}{n_{\mathrm{pf}}^{2} \cdot \left(S_{\mathrm{PID}} + 0.94\right)^{2}} \right)}$&0.00011&0.849&64.02&0.93\\
14&$0.96 \cdot \tanh^{3}{\left(0.3 + \frac{0.65}{n_{\mathrm{pf}}^{2} \cdot \left(S_{\mathrm{PID}} + 0.85\right)^{2}} \right)}$&0.00009&0.849&64.29&0.83\\
17&$\left(- 0.032 \cdot\sqrt{ n_{\mathrm{pf}}} + \tanh{\left(0.32 + \frac{0.64}{n_{\mathrm{pf}}^{2} \cdot \left(S_{\mathrm{PID}} + 0.82\right)^{2}} \right)}\right)^{3}$&0.00008&0.849&63.94&0.85\\
18&$\left(- 0.032 \times \tanh^{0.5}{\left(n_{\mathrm{pf}} \right)} + \tanh{\left(0.32 + \frac{0.64}{n_{\mathrm{pf}}^{2} \times \left(S_{\mathrm{PID}} + 0.81\right)^{2}} \right)}\right)^{3}$&0.00008&0.849&63.94&0.82\\
20&$\left(- 0.032 \cdot \tanh^{0.5}{\left(1.06 \cdot n_{\mathrm{pf}} \right)} + \tanh{\left(0.32 + \frac{0.64}{n_{\mathrm{pf}}^{2} \times \left(S_{\mathrm{PID}} + 0.82\right)^{2}} \right)}\right)^{3}$&0.00007&0.849&60.85&0.86\\
\bottomrule
\end{tabular}
\end{scriptsize}
\end{small}
\caption{2D symbolic regression tables for $\npf$ and $\spid$}
\end{table}
\begin{table}
\centering
\begin{small}
\begin{scriptsize}
\begin{tabular}{c|p{9cm}|S[table-format=1.5]|S[table-format=1.3]|S[table-format=2.2]|S[table-format=2.2]}
\toprule
 Complexity& Formula& {Loss}& {AUC} &{$\rej_{30\%}$}& {$\Delta C$} \\
 \midrule
1&$\ptd$&0.07802&0.807&26.87&21.19\\
2&$\ptd^{0.5}$&0.05645&0.807&26.87&15.52\\
3&$0.93 - \npf$&0.02289&0.840&52.32&11.05\\
4&$1.7 \cdot \left(1.0 - 0.84 \cdot \npf\right)^{3}$&0.00592&0.840&52.32&4.13\\
5&$\tanh{\left(\frac{0.08}{\npf^{2}} \right)}$&0.00406&0.840&52.32&3.42\\
6&$- \tanh{\left(0.22- \frac{\ptd}{\npf} \right)}$&0.00317&0.842&52.14&2.41\\
7&$\tanh{\left(\frac{0.26\cdot \ptd}{\npf^{2}} \right)}$&0.00136&0.844&55.01&1.02\\
8&$\tanh^{3}{\left(\frac{\ptd^{2} + 0.32}{\npf} \right)}$&0.00111&0.844&54.00&1.48\\
9&$\tanh{\left(\left(\ptd - 0.27 + \frac{0.27}{\npf}\right)^{2} \right)}$&0.00093&0.845&54.70&1.39\\
10&$0.94 \cdot \tanh{\left(\left(\ptd + \frac{0.068}{\npf^{2}}\right)^{2} \right)}$&0.00076&0.845&54.00&0.73\\
11&$\tanh^{2}{\left(\left(\ptd + 0.35\right)^{3} + \frac{0.08}{\npf^{2}} \right)}$&0.00070&0.845&54.17&1.18\\
12&$0.95 \cdot \tanh{\left(4.65\cdot \ptd^{3} + \frac{0.03}{\npf^{3}} \right)}$&0.00044&0.845&53.94&0.64\\
13&$0.95 \cdot \tanh{\left(\left(\ptd + 0.41\right)^{6} + \frac{0.025}{\npf^{3}} \right)}$&0.00044&0.845&53.88&0.94\\
14&$0.95 \cdot \tanh{\left(4.41 \cdot \ptd^{3} + 0.01 + \frac{0.02}{\npf^{3}} \right)}$&0.00042&0.845&53.71&0.60\\
16&$0.95 \cdot \tanh{\left(6.16\cdot \left(0.21 - \ptd\right)^{2} + 0.02 \cdot \left(0.49 + \frac{1}{\npf}\right)^{3} \right)}$&0.00037&0.846&54.23&0.58\\
17&$0.95 \cdot \tanh{\left(5.28 \cdot \ptd^{3} + \frac{0.02}{\left(\npf - \frac{0.0005}{\ptd^{3}}\right)^{3}} \right)}$&0.00031&0.846&53.13&1.07\\
18&$0.95 \cdot \tanh{\left(0.019\cdot \left(0.41 + \frac{1}{\npf}\right)^{3} + \left(\ptd \cdot \left(\npf + 2.30\right) - 0.63\right)^{2} \right)}$&0.00031&0.846&53.82&0.64\\
19&$\tanh{\left(5.59\cdot \ptd^{3} + 4836.5 \cdot \left(\left(0.37 - \ptd\right)^{3} + \frac{0.017}{\npf}\right)^{3} \right)} - 0.047$&0.00025&0.846&53.76&0.80\\
21&$0.95 \cdot \tanh{\left(5.79\cdot \ptd^{3} + 0.03 \cdot \left(-0.23 + \frac{1}{\npf + 9.96 \cdot \left(\ptd - 0.39\right)^{3}}\right)^{3} \right)}$&0.00022&0.846&52.63&0.68\\
\bottomrule
\end{tabular}
\end{scriptsize}
\end{small}
\caption{2D symbolic regression tables for $\npf$ and $\ptd$}
\end{table}

\begin{table}
\centering
\begin{small}
\begin{scriptsize}
\begin{tabular}{c|p{9cm}|S[table-format=1.5]|S[table-format=1.3]|S[table-format=2.2]|S[table-format=2.2]}
\toprule
 Complexity& Formula& {Loss}& {AUC} &{$\rej_{30\%}$}& {$\Delta C$} \\
 \midrule
1&0.50&0.10417&0.5&3.33& {-}\\
3&$0.94 - \npf$&0.03430&0.840&52.32&11.12\\
4&$1.7 \cdot \left(1.0 - 0.84 \cdot \npf\right)^{3}$&0.01592&0.840&52.32&3.90\\
5&$\tanh{\left(\frac{0.03}{\npf^{3}} \right)}$&0.01307&0.840&52.32&2.90\\
6&$\tanh^{6}{\left(\frac{0.57}{\npf} \right)}$&0.01262&0.840&52.32&2.05\\
7&$\tanh{\left(\frac{0.04}{\npf^{3} + 0.01} \right)}$&0.01226&0.840&52.32&1.15\\
8&$\tanh{\left(\frac{0.31 \cdot \sqrt{\frac{1}{r_\lambda}}}{\npf^{3}} \right)}$&0.00612&0.852&46.04&2.04\\
9&$\tanh^{2}{\left(\frac{1.28 \cdot \sqrt{\frac{1}{r_\lambda}}}{\npf^{2}} \right)}$&0.00470&0.852&41.25&2.37\\
10&$\tanh^{9}{\left(\frac{189.7}{\npf \cdot \left(r_\lambda + 198.6\right)} \right)}$&0.00391&0.855&42.84&2.06\\
11&$\tanh^{3}{\left(9.27 \cdot \left(\sqrt{\frac{1}{r_\lambda}} + \frac{0.091}{\npf}\right)^{2} \right)}$&0.00337&0.856&43.52&1.90\\
12&$\tanh^{9}{\left(7.32 \cdot \sqrt{\frac{1}{r_\lambda}} + \frac{0.13}{\npf^{2}} \right)}$&0.00238&0.858&46.00&1.47\\
13&$\tanh^{9}{\left(\frac{123.8}{r_\lambda + 68.67} + \frac{0.13}{\npf^{2}} \right)}$&0.00190&0.859&48.59&1.64\\
14&$\tanh^{18}{\left(\frac{236.17}{r_\lambda + 120.8} + \frac{0.13}{\npf^{2}} \right)}$&0.00184&0.859&49.36&1.68\\
15&$\left(1.1 - \npf\right) \cdot \tanh{\left(181.1 \cdot \left(\sqrt{\frac{1}{r_\lambda}} + \frac{0.003}{\npf^{3}}\right)^{3} \right)}$&0.00115&0.860&51.81&1.41\\
16&$\left(1.1 - \npf\right) \cdot \tanh^{3}{\left(54.53\cdot \left(\sqrt{\frac{1}{r_\lambda}} + \frac{0.003}{\npf^{3}}\right)^{2} \right)}$&0.00083&0.860&51.81&1.43\\
17&$\left(1.1 - \npf\right) \cdot \tanh^{1.5}{\left(268.4\cdot \left(\sqrt{\frac{1}{r_\lambda}} + \frac{0.003}{\npf^{3}}\right)^{3} \right)}$&0.00082&0.860&51.81&1.77\\
18&$\tanh{\left(\frac{0.36}{\npf} \right)} \cdot \tanh{\left(464.46 \cdot \left(\sqrt{\frac{1}{r_\lambda}} - 0.05 + \frac{0.01}{\npf^{2}}\right)^{3} \right)}$&0.00078&0.860&51.81&1.59\\
19&$\tanh{\left(\frac{0.36}{\npf} \right)} \cdot \tanh{\left(1273.42 \cdot \left(0.01\cdot \left(-0.60 + \frac{1}{\npf}\right)^{2} + \sqrt{\frac{1}{r_\lambda}}\right)^{4} \right)}$&0.00076&0.860&51.81&1.51\\
20&$\tanh{\left(\frac{0.36}{\npf} \right)} \cdot \tanh{\left(464.25 \cdot \left(\sqrt{\frac{1}{r_\lambda}} - 0.04 + \frac{0.007}{\left(\npf - 0.05\right)^{2}}\right)^{3} \right)}$&0.00075&0.860&51.81&1.52\\
21&$\left(- \npf - 0.59 \cdot \left(\frac{1}{r_\lambda}\right)^{0.5} + 1.2\right) \cdot \tanh{\left(1472.86 \cdot \left(\sqrt{\frac{1}{r_\lambda}} + \frac{0.0026}{\npf^{3}}\right)^{4} \right)}$&0.00069&0.860&54.41&1.60\\
22&$\left(- \npf - 0.57 \cdot \left(\frac{1}{r_\lambda}\right)^{0.5} + 1.2\right) \cdot \left(\tanh^{3}{\left(272.2 \cdot \left(\sqrt{\frac{1}{r_\lambda}} + \frac{0.003}{\npf^{3}}\right)^{3} \right)}\right)^{0.5}$&0.00065&0.860&54.76&1.46\\
\bottomrule
\end{tabular}
\end{scriptsize}
\end{small}

\caption{2D symbolic regression tables for $\npf$ and $r_\lambda$}
\end{table}

\begin{table}

\centering
\begin{small}
\begin{scriptsize}
\begin{tabular}{c|p{9cm}|S[table-format=1.5]|S[table-format=1.3]|S[table-format=2.2]|S[table-format=2.2]}
\toprule
 Complexity& Formula& {Loss}& {AUC} &{$\rej_{30\%}$}& {$\Delta C$} \\
 \midrule
1&$0.5$&0.08389&0.500&3.33 & {-} \\
3&$0.93 - \npf$&0.01909&0.840&52.32&11.04\\
4&$1.7 \cdot \left(1.0 - 0.84 \cdot \npf\right)^{3}$&0.00358&0.840&52.32&4.14\\
5&$\tanh{\left(\frac{0.08}{\npf^{2}} \right)}$&0.00156&0.840&52.32&3.53\\
6&$\tanh^{6}{\left(\frac{0.57}{\npf} \right)}$&0.00113&0.840&52.32&2.01\\
7&$\tanh{\left(\frac{0.04}{\npf^{3} + 0.01} \right)}$&0.00069&0.840&52.32&1.22\\
9&$0.93 \cdot \tanh{\left(\frac{0.055}{\left(\npf - 0.086\right)^{2}} \right)}$&0.00053&0.830&52.32&0.96\\
10&$\tanh{\left(\frac{0.04}{\npf^{3} + 0.02 \cdot \sqrt{E_Q}} \right)}$&0.00047&0.840&52.36&1.29\\
11&$- 0.098 \cdot E_Q + \tanh{\left(0.11 + \frac{0.03}{\npf^{3}} \right)}$&0.00040&0.840&52.36&0.84\\
12&$0.94 \cdot \tanh{\left(\frac{0.06}{\left(\npf + 0.08 \cdot \sqrt{E_Q}\right)^{3}} \right)}$&0.00032&0.840&52.41&0.86\\
13&$0.9 \cdot \tanh{\left(\frac{0.04}{\left(\npf + 0.05 \cdot E_Q\right)^{3}} \right)} + 0.04$&0.00026&0.840&52.69&0.88\\
14&$0.92 \cdot \tanh{\left(\frac{0.05}{\left(\npf + 0.06 \cdot \sqrt{E_Q}\right)^{3}} \right)} + 0.027$&0.00025&0.840&52.52&0.94\\
16&$0.18 - 0.76 \cdot \tanh{\left(0.19 - \frac{0.06}{\left(\npf + 0.07 \cdot \sqrt{E_Q}\right)^{3}} \right)}$&0.00023&0.840&52.36&0.85\\
17&$0.32 - 0.62 \cdot \tanh{\left(0.49 - \frac{0.098}{\left(\npf + 0.13 \cdot \sqrt[4]{E_Q}\right)^{3}} \right)}$&0.00022&0.840&52.52&0.87\\
18&$0.42 - 0.52 \cdot \tanh{\left(0.92 - \frac{0.24}{\left(\npf + 0.23\cdot \sqrt[8]{E_Q}\right)^{3}} \right)}$&0.00021&0.840&52.36&0.94\\
19&$0.18 - 0.76 \cdot \tanh{\left(0.19 - \frac{0.06}{\left(\npf + E_Q \cdot \left(0.18 - 0.11 \cdot E_Q\right)\right)^{3}} \right)}$&0.00021&0.840&52.69&0.92\\
20&$0.36 - 0.58 \cdot \tanh{\left(0.62 - \frac{0.13}{\left(\npf + 0.32 \cdot \sqrt{E_Q} - 0.17\cdot E_Q\right)^{3}} \right)}$&0.00020&0.840&52.52&0.92\\
21&$0.91 \cdot \tanh{\left(\frac{0.05}{\left(\npf + \frac{0.09\cdot E_Q}{\sqrt{\left(\npf - E_Q\right)^{2}} + 0.58}\right)^{3}} \right)} + 0.034$&0.00017&0.840&52.08&0.99\\
22&$0.91 \cdot \tanh{\left(\frac{0.05}{\left(\npf + \frac{0.07\cdot E_Q}{\sqrt{\sqrt{\left(\npf - E_Q\right)^{2}} + 0.10}}\right)^{3}} \right)} + 0.034$&0.00017&0.840&51.92&0.95\\

\bottomrule
\end{tabular}
\end{scriptsize}
\end{small}

\caption{2D symbolic regression tables for $\npf$ and $E_Q$}
\end{table}

\begin{table}
\centering
\begin{scriptsize}
\begin{tabular}{c|l|S[table-format=1.5]|S[table-format=1.3]|S[table-format=2.2]|S[table-format=2.2]}
\toprule
 Complexity& Formula& {Loss}& {AUC} &{$\rej_{30\%}$}& {$\Delta C$} \\
 \midrule
1&$0.49$&0.09075&0.5&3.33& {-}\\
3&$0.92 - \npf$&0.02469&0.840&52.32&11.35\\
4&$1.7 \cdot \left(1.0 - 0.84 \cdot \npf\right)^{3}$&0.00795&0.840&52.32&4.25\\
5&$\tanh{\left(\frac{0.082}{\npf^{2}} \right)}$&0.00612&0.840&52.32&3.59\\
6&$\tanh^{6}{\left(\frac{0.56}{\npf} \right)}$&0.00532&0.840&52.32&2.05\\
7&$\tanh{\left(\frac{0.06}{\left(\npf + 0.10\right)^{3}} \right)}$&0.00510&0.840&52.32&1.49\\
8&$\tanh{\left(\frac{0.10}{\left(\npf + 0.26\right)^{4}} \right)}$&0.00506&0.840&52.32&1.49\\
9&$0.94 \cdot \tanh{\left(0.02 \cdot \left(0.37 + \frac{1}{\npf}\right)^{3} \right)}$&0.00484&0.840&52.32&1.08\\
10&$\tanh{\left(\frac{0.07}{\left(\npf - \left(C_{0.2} - 0.58\right)^{2}\right)^{2}} \right)}$&0.00444&0.841&61.58&2.00\\
11&$\tanh^{4}{\left(\frac{0.46}{- \npf + \left(C_{0.2} - 0.57\right)^{2}} \right)}$&0.00417&0.841&60.46&1.32\\
12&$- \tanh{\left(C_{0.2} - \frac{\sqrt{\left(C_{0.2} - 0.46\right)^{2}} + 0.35}{\npf} \right)}$&0.00265&0.846&62.81&2.43\\
13&$\tanh^{2}{\left(C_{0.2} - \frac{\sqrt{\left(C_{0.2} - 0.447\right)^{2}} + 0.48}{\npf} \right)}$&0.00233&0.846&61.5&1.55\\
14&$\tanh^{3}{\left(- C_{0.2} + \sqrt[4]{\left(C_{0.2} - 0.45\right)^{2}} + \frac{0.53}{\npf} \right)}$&0.00199&0.846&59.03&1.34\\
15&$\tanh^{2}{\left(- \sqrt{C_{0.2}} + \sqrt[4]{\left(C_{0.2} - 0.45\right)^{2}} + \frac{0.53}{\npf} \right)}$&0.00181&0.846&59.03&1.36\\
16&$\tanh^{6}{\left(131.67 \cdot \left(\left(C_{0.2} - 0.68\right)^{2} - 0.05\right)^{2} + \frac{0.52}{\npf} \right)}$&0.00139&0.847&64.02&1.88\\
17&$\tanh^{3}{\left(- C_{0.2} + 3.88\cdot \sqrt{\left(\left(0.71 \cdot C_{0.2} - 1\right)^{4} - 0.21\right)^{2}} + \frac{0.54}{\npf} \right)}$&0.00122&0.848&62.34&1.61\\
18&$\left(0.03 - \tanh^{3}{\left(\frac{82.47\cdot \left(\left(C_{0.2} - 0.68\right)^{2} - 0.04\right)^{2} + 0.54}{\npf} \right)}\right)^{2}$&0.00089&0.848&65.27&1.30\\
19&$\left(\tanh{\left(- C_{0.2} + 287 \cdot \left(\left(0.70 \cdot C_{0.2} - 1\right)^{8} - 0.05\right)^{2} + \frac{0.58}{\npf} \right)} - 0.019\right)^{3}$&0.00078&0.848&65.10&1.51\\
21&$\tanh^{3}{\left(- C_{0.2} + 288.3\cdot \left(\left(0.70\cdot C_{0.2} - 1\right)^{8} - 0.05\right)^{2} + \frac{0.64}{\npf + 0.03} \right)} - 0.05$&0.00074&0.848&65.19&1.33\\
22&$\left(\tanh^{3}{\left(- C_{0.2} + 300.9 \cdot \left(\left(0.70 \cdot C_{0.2} - 1\right)^{8} - 0.05\right)^{2} + \frac{0.89}{\npf + 0.09} \right)} - 0.027\right)^{2}$&0.00073&0.848&64.77&1.18\\
\bottomrule
\end{tabular}
\end{scriptsize}
\caption{2D symbolic regression tables for $\npf$ and $C_{0.2}$}
\end{table}

\begin{table}
\centering
\begin{scriptsize}
\begin{tabular}{c|p{9cm}|S[table-format=1.5]|S[table-format=1.3]|S[table-format=2.2]|S[table-format=2.2]}
\toprule
 Complexity& Formula& {Loss}& {AUC} &{$\rej_{30\%}$}& {$\Delta C$} \\
 \midrule
1&$0.5$&0.09448&0.5&3.33\\
3&$0.93 - \npf$&0.02494&0.840&52.32&11.07\\
4&$1.7 \cdot \left(1.0 - 0.84 \cdot \npf\right)^{3}$&0.00669&0.840&52.32&4.06\\
5&$\tanh{\left(\frac{0.03}{\npf^{3}} \right)}$&0.00372&0.840&52.32&2.78\\
6&$\tanh^{6}{\left(\frac{0.57}{\npf} \right)}$&0.00327&0.840&52.32&2.06\\
7&$\tanh^{3}{\left(\frac{1.18}{\npf \cdot \sfrag} \right)}$&0.00151&0.844&54.35&1.74\\
9&$0.94 \cdot \tanh{\left(\frac{0.69}{\npf^{2} \cdot \sfrag^{2}} \right)}$&0.00109&0.844&54.29&1.32\\
11&$0.95 \cdot \tanh^{2}{\left(\frac{0.73}{\sfrag \cdot \left(\npf - 0.11\right)} \right)}$&0.00099&0.844&55.80&1.03\\
12&$0.93 \cdot \tanh{\left(1.89 \cdot \left(- \frac{1}{\sfrag} - \frac{0.15}{\npf}\right)^{4} \right)}$&0.00096&0.844&55.80&1.33\\
13&$0.94 \cdot \tanh^{2}{\left(\frac{0.57}{\left(\npf - 0.13\right) \cdot \left(\npf - \sfrag\right)} \right)}$&0.00092&0.844&55.74&0.93\\
14&$0.94 \cdot \tanh{\left(0.96 \cdot \left(-0.06- \frac{1}{\sfrag} - \frac{0.18}{\npf}\right)^{4} \right)}$&0.00089&0.844&55.31&0.95\\
15&$0.95 \cdot \tanh^{2}{\left(0.28\cdot \left(-0.52 - \frac{1}{\sfrag} - \frac{0.18}{\npf}\right)^{4} \right)}$&0.00086&0.844&55.37&0.94\\
16&$0.93 \cdot \tanh{\left(\frac{\left(\left(\npf - 0.17\cdot \sfrag\right)^{2} + \frac{0.81}{\sfrag}\right)^{2}}{\npf^{2}} \right)}$&0.00082&0.844&55.31&1.60\\
17&$0.94 \cdot \tanh{\left(\frac{\left(\left(- \npf + 0.059\cdot \sfrag^{2}\right)^{2} + \frac{0.81}{\sfrag}\right)^{2}}{\npf^{2}} \right)}$&0.00045&0.845&54.53&1.11\\
18&$0.94 \cdot \tanh{\left(\frac{\left(\sqrt{\left(- \npf + 0.06 \cdot \sfrag^{2}\right)^{2}} + 0.77\right)^{2}}{\npf^{2} \cdot \sfrag^{2}} \right)}$&0.00040&0.845&53.59&1.15\\
19&$0.94 \cdot \tanh{\left(\frac{4\cdot \left(\left(\npf - 0.06 \cdot \sfrag^{2}\right)^{2} + \frac{0.40}{\sfrag}\right)^{2}}{\npf^{2}} \right)}$&0.00030&0.845&53.88&1.09\\
21&$0.93 \cdot \tanh{\left(\frac{3.90 \cdot \left(\left(- \npf + 0.06 \cdot \sfrag^{2}\right)^{2} + \frac{0.40}{\sfrag}\right)^{2}}{\npf^{2}} \right)} + 0.0092$&0.00029&0.845&53.88&0.89\\
22&$0.94 \cdot \tanh^{2}{\left(2.74\cdot \left(0.19 + \frac{\left(- \npf + 0.05\cdot \sfrag^{2}\right)^{2} + \frac{0.40}{\sfrag}}{\npf}\right)^{2} \right)}$&0.00027&0.845&53.88&1.03\\

\bottomrule
\end{tabular}
\end{scriptsize}
\caption{2D symbolic regression tables for $\npf$ and $\sfrag$}
\end{table}

\clearpage
\bibliography{tilman,refs}
\end{document}